\documentclass[sigconf,screen]{acmart}

\usepackage{comment}
\usepackage{soul}
\usepackage{enumitem}
\usepackage{xcolor}
\usepackage{framed}

\setcopyright{acmcopyright}
\copyrightyear{2025}
\acmYear{2025}
\acmDOI{XXXXXXX.XXXXXXX}

\acmConference[arXiv]{arXiv}{April}{2025}
\newcommand{\sys}{\textsc{DimInd}}

\begin{document}

\settopmatter{printacmref=false,printccs=false}
\setcopyright{none}
\renewcommand\footnotetextcopyrightpermission[1]{}

\title{Facets, Taxonomies, and Syntheses: Navigating Structured Representations in LLM-Assisted Literature Review}

\author{Raymond Fok}
\email{rayfok@cs.washington.edu}
\affiliation{
  \institution{University of Washington}
  \city{Seattle}
  \state{WA}
  \country{USA}
}

\author{Joseph Chee Chang}
\email{josephc@allenai.org}
\affiliation{
  \institution{Allen Institute for AI}
  \city{Seattle}
  \state{WA}
  \country{USA}
}

\author{Marissa Radensky}
\email{radensky@cs.washington.edu}
\affiliation{
  \institution{University of Washington}
  \city{Seattle}
  \state{WA}
  \country{USA}
}

\author{Pao Siangliulue}
\email{paos@allenai.org}
\affiliation{
  \institution{Allen Institute for AI}
  \city{Seattle}
  \state{WA}
  \country{USA}
}

\author{Jonathan Bragg}
\email{jbragg@allenai.org}
\affiliation{
  \institution{Allen Institute for AI}
  \city{Seattle}
  \state{WA}
  \country{USA}
}

\author{Amy X. Zhang}
\email{axz@cs.uw.edu}
\affiliation{
  \institution{University of Washington}
  \city{Seattle}
  \state{WA}
  \country{USA}
}

\author{Daniel S. Weld}
\email{danw@allenai.org}
\affiliation{
  \institution{Allen Institute for AI \&\ \\ University of Washington}
  \city{Seattle}
  \state{WA}
  \country{USA}
}



\begin{teaserfigure}
    \centering
    \includegraphics[width=\textwidth]{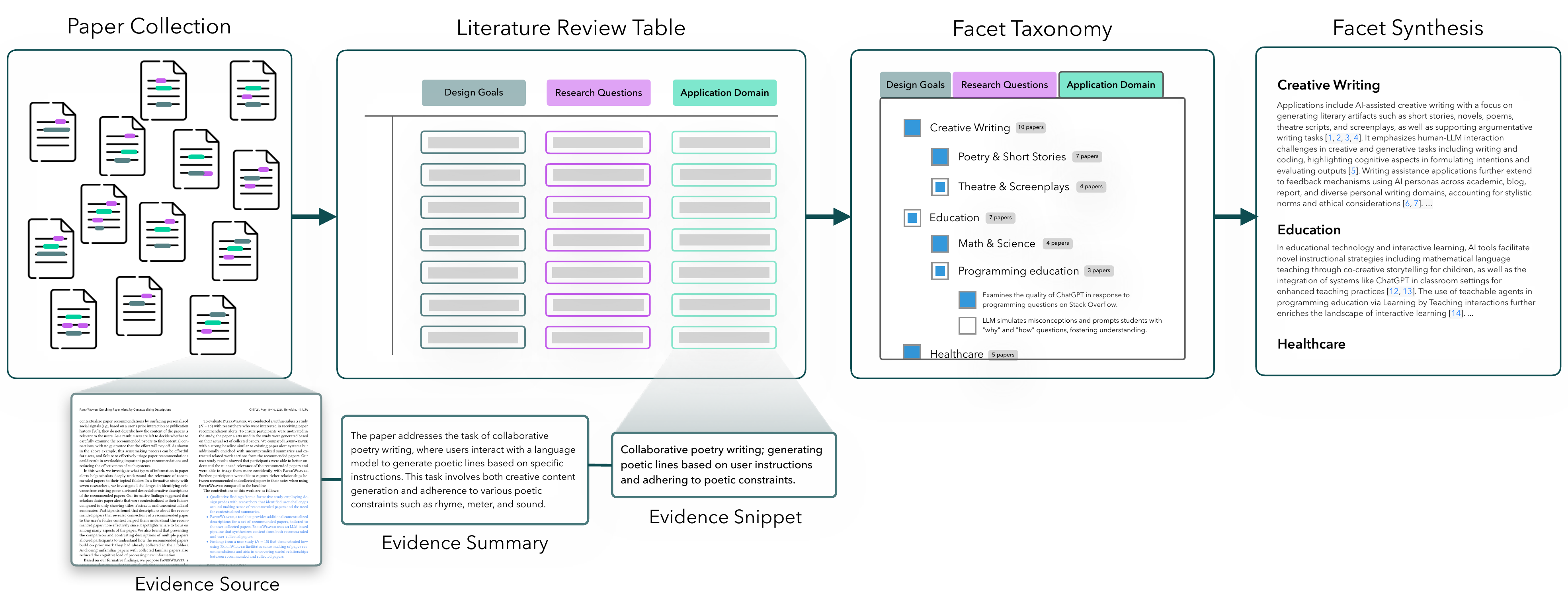}
    \caption{We present an LLM-assisted workflow aimed at scaffolding literature review over large paper collections, and instantiate it in a prototype system, \sys{}. Users can interactively construct and explore four successive \textit{structured representations} of literature information: a \textit{paper collection} listing papers and metadata, a \textit{literature review table} with columns that render relevant evidence snippets from paper full texts along defined facets, a \textit{facet taxonomy} that organizes faceted information into a higher-level conceptual overview, and a \textit{facet synthesis} that provides a controllable faceted summary across the collection.}
    \Description{}
    \label{fig:teaser}
\end{teaserfigure}

\begin{abstract}
    Comprehensive literature review requires synthesizing vast amounts of research---a labor intensive and cognitively demanding process. Most prior work focuses either on helping researchers deeply understand a few papers (e.g., for triaging or reading), or retrieving from and visualizing a vast corpus. Deep analysis and synthesis of large paper collections (e.g., to produce a survey paper) is largely conducted manually with little support. We present \sys, an interactive system that scaffolds literature review across large paper collections through LLM-generated structured representations. \sys{} scaffolds literature understanding with multiple levels of compression, from papers, to faceted literature comparison tables with information extracted from individual papers, to taxonomies of concepts, to narrative syntheses. Users are guided through these successive information transformations while maintaining provenance to source text. In an evaluation with 23 researchers, \sys{} supported participants in extracting information and conceptually organizing papers with less effort compared to a ChatGPT-assisted baseline workflow.
\end{abstract}

\maketitle

\section{Introduction}
Literature review offers a foundation for scientific progress.
Researchers spend significant effort making sense of prior work throughout a research project, from finding inspirations for ideation, exploring techniques and datasets to develop methods, to writing related work to contextualize their own work. 
One way this effort can be shared and reused is in the form of survey papers that comprehensively review a large collection of papers.
However, the process of reviewing and synthesizing a large set of papers is time-intensive~\cite{michelson2019significantCost, shojania2007quickly, borah2017analysis} and cognitively challenging~\cite{knight2019enslaved, granello2001promoting}.
As a result, this process is often conducted with a team of researchers over multiple weeks or months and rarely feasible for individual scholars.
For example,~\citet{pang2025llmification} had seven authors who collectively reviewed 186 papers, while \citet{lee2024writingassistants} had an even bigger team of 36 researchers reviewing 274 papers.
In the field of medicine, Cochrane coordinates a team of over 28,000 contributors to produce its timely review articles~\cite{Chandler2013CochraneM}.

One main challenge is the sheer amount of information scattered across disparate papers that needs to be extracted, analyzed, compared, and synthesized \cite{snyder2019literature,lee2024writingassistants}.
As a result, researchers often follow a step-by-step process to make the tasks more tractable, involving successively transforming information from one representation to another that is one degree more synthesized.
For instance, the process begins with gathering a large collection of relevant papers, then reading and summarizing those papers individually, coding the summaries into common themes or aspects in large spreadsheets~\cite{hashimoto2017automatic, newman2024arxivdigestables}, building out a taxonomy of research threads~\cite{kang_threddy_2022, palani_relatedly_2023}, and finally, synthesizing this structure into writing.
Information management and coordination often involves large tabular spreadsheets to keep track of papers, summaries, qualitative codes, and themes.\footnote{e.g., \url{https://writing-assistant.github.io/\#annotated-papers}, \url{https://github.com/Social-Futures-Lab/skin-deep/blob/main/Literature\%20Review.csv}
}
Currently, much of this work is labor-intensive and low-level, involving significant back-and-forth---from manually inspecting the full text of papers and extracting information across many facets, to revisiting the full text to check for accuracy---incurring substantial cognitive costs~\cite{pirolli2005sensemaking, cooper2015research, khalil2022challenges}. 
And rapid growth in scholarly publication means that the effort required to produce such reviews will only continue to increase~\cite{bornmann2015growth, jinha2010article}.

Recent work has suggested large language models (LLMs) can be effective at facilitating this information extraction and compression process over large paper collections but these explorations have largely focused on \emph{``data tables''} or task-specific structures, such as size and accuracy of AI models or intervention and patient demographic in clinical trials~\cite{elicit, wang2024scidasynth}.
In contrast, their ability to extract nuanced qualitative insights that are useful for researchers to create more narrative literature reviews remains underexplored, and more importantly, existing approaches typically produce only flat, single-level representations (such as tables) rather than the multiple, interconnected levels of abstraction needed to support the progressive synthesis of comprehensive literature review.

In this work, we present \sys{}, an interactive system for LLM-powered literature review support that guides researchers through successive structured representations of paper information.
Starting with a \textbf{collection of papers}, users iteratively construct a faceted \textbf{literature review table} that organizes relevant information extracted from papers' full texts by defining custom facets in natural language or selecting from collection-aware facets suggested by the system.
The system then transforms each faceted column into a hierarchical \textbf{taxonomy of concepts}, surfacing emergent themes across papers.
These taxonomies can be then explored, refined, and transformed into \textbf{narrative syntheses}.
Users are guided through these structured representations via interaction and visual affordances.
This progressive disclosure not only scaffolds sensemaking, but also supports the steering and verification of information, which is critical when working with LLM-generated content.

To evaluate how \sys{} can better support analyzing and synthesizing large collections of papers, we conducted a within-subjects user evaluation with 23 computer science researchers where they reviewed two sets of 50 research papers in two 30-minute sessions.
Comparing \sys{} against a baseline literature review approach assisted by a commercial LLM-based chat application (ChatGPT), we found that \sys{}'s structured representations effectively supported users in extracting, organizing, and verifying information across papers with less effort.
Our qualitative findings further reveal how researchers used tables as information scent and taxonomies as navigational hubs, transforming information into manageable views that facilitate movement between high-level organization and detailed exploration, while balancing LLM assistance with their own scholarly agency.
We conclude by discussing the tradeoffs of structured versus conversational assistance and highlight opportunities and tensions in LLM-assisted literature review workflows. In sum, this paper contributes:
\begin{itemize}
    \item Design goals for an LLM-assisted workflow for literature review support, informed by sensemaking theories and prior work, that address the tedium and cognitive challenges in extracting, organizing, and making sense of information across a large body of literature through successive structured representations.
    \item \sys{}, an interactive system that instantiates these ideas, with features to support users in exploring and navigating between transformations of literature information.
    \item Findings from a within-subjects user evaluation with 23 researchers, demonstrating that \sys{} reduced cognitive load and increased paper organization effectiveness during exploratory literature review compared to traditional methods, with qualitative insights into how participants use and navigate across the various structured representations.
\end{itemize}

\section{Related Work}

\subsection{Making Sense of Unstructured Data}

Literature review is an example of a \textit{sensemaking} process in which researchers collect, extract, transform, and synthesize scattered scholarly information into coherent structures~\cite{russell_cost_1993}. 
This process of creating and refining schemas in literature review closely parallels practices in qualitative analysis~\cite{corbin1990grounded, braun_thematicAnalysis_2006}. Both involve identifying relevant segments from unstructured data (e.g., quotes from a transcript or snippets from a document), identifying patterns and themes, and creating structures to keep track of them (e.g., codes, tables, taxonomies). For qualitative analysis, early approaches like topic modeling can help analyze large text collections~\cite{blei2003latent, griffiths2004finding}; however these methods often produced unsatisfactory analyses---irrelevant \cite{chang2009reading, alsumait_topic_2009}, uninterpretable~\cite{doogan_topic_2021}, or misaligned~\cite{hoyle_neural_2022}.

More recent work has explored leveraging large language models (LLMs) for automating qualitative analysis, revisiting tasks such as topic modeling~\cite{pham_topicgpt_2023, wang2023prompting} and document clustering~\cite{zhang_clusterllm_2023, wang2023goaldriven} to greater success. 
Inspired by these successes, we explore if and how LLMs can similarly induce higher-order structures from large amounts of complex scholarly information toward scaffolding an interactive literature review workflow.

Relatedly, prior work has explored mixed-initiative systems \cite{horvitz1999principles} for general qualitative data analysis. Tools like Cody~\cite{rietz_cody_2021}, Scholastic~\cite{hong_scholastic_2022}, and CoAIcoder~\cite{gao2023coaicoder} demonstrate how AI can support qualitative researchers in organizing and analyzing text, while systems like LLooM~\cite{lam2024concept} extract high-level facets to guide initial directions of analysis.
In this paper, we focus on applying mixed-initiative principles to literature review, presenting interactive structured representations---faceted tables, taxonomies, and narratives---that scaffold the foraging and sensemaking activities of literature review while keeping researchers connected to the underlying papers.

\subsection{Deep Engagement with Small Sets of Papers} \label{sec:small_set}
With the exponential growth of scientific publication and recent advancement in LLMs, the HCI research community has become increasingly excited about building \emph{user-driven tools} that can better support understanding the literature at various scales.

One line of work has explored augmented reading interfaces to reduce cognitive load when engaging with \emph{individual papers}. These include tools that aid comprehension by simplifying scholarly language~\cite{head_augmenting_2021, august2023paperPlain}, improve efficiency by selectively guiding reader attention~\cite{fok_scim_2023, han2022passages}, and lower the costs of saving relevant information~\cite{han2022passages, kang_threddy_2022}. Other tools weave external context into the paper environment, for example by visually augmenting inline citations based on a user's reading history~\cite{chang_citesee_2023}, or embedding relevant follow-on work~\cite{rachatasumrit_citeread_2022} and presentation videos~\cite{kim2023papeos} as margin-notes localized to relevant parts of a paper.

Beyond reading support for individual papers, prior work has also explored how to help researchers more deeply understand a small set of papers. For example, systems such as PaperWeaver~\cite{lee2024paperweaver} or ACCoRD~\cite{murthy2022accord} generate comparative statements between of \emph{two papers or concepts} to help researchers better understand an unfamiliar paper by contextualizing it to a familiar paper. More closely related to our work, systems like Elicit~\cite{elicit} and SciDaSynth~\cite{wang2024scidasynth} allow users to compare small sets of papers in a fixed table-based representation. Tables are ubiquitous across diverse sensemaking domains, such as online search~\cite{chang_searchlens_2019, spenke_focus_1996, chang_mesh_2020}, developer support~\cite{liu_unakite_2019, liu_crystalline_2022}, and business analysis~\cite{fok2024marco}; within scholarly sensemaking, researchers also often manually create literature review tables to organize, compare, and synthesize information across many papers~\cite{hashimoto2017automatic, newman2024arxivdigestables, wang2024scidasynth}. However, at the scale of larger paper collections (e.g., dozens to hundreds), tables can become insufficient for sensemaking, and require complementary affordances and flexible representations to provide additional cognitive support.

\subsection{Understanding Literature at Scale} \label{sec:large_set}
While there are also user-driven systems that help users engage with papers at scale, these have focused on supporting \emph{searching and navigation} along high-level topics. For example, prior approaches cluster a paper's references in a reading environment into related research threads for exploration~\cite{kang_synergi_2023, kang_threddy_2022}, visually cluster papers on a canvas based on citation-edges or semantic similarity~\cite{Narechania_2022,connectedpapers}, organize papers in hierarchical structures~\cite{hsu2024chime, zhu2023hierarchical}, and create overarching narrative summaries~\cite{susnjak2025automating, altmami2022automatic}.
While these approaches may be effective for broadly searching and navigating over the literature landscape, they lack support for deeper comparison and synthesis over a large collection of papers. They often stop short of supporting the schematization aspect of sensemaking, particularly in organizing nuanced relationships grounded in information from papers' full texts. 

Another major thread of research has focused on fully automating literature review and survey paper writing~\cite{susnjak2025automating, altmami2022automatic, erera2019summarization, lu2020multixscience}. These techniques are promising but nascent~\cite{qureshi2023chatgpt, bolanos2024artificial}, facing challenges in the quality of generated syntheses~\cite{martinboyle2024shallow, deyoung2023multi}, hallucination~\cite{george2023factoredverification, belem2024singlemultillmshallucinate}, and alignment with researchers' specific goals~\cite{qureshi2023chatgpt, yang2023oassum}.
More fundamentally, fully automated methods still struggle to capture the nuanced and iterative nature of sensemaking, including both qualitative analysis and literature review. Analysts often want to ``stay close to the data,'' and are skeptical of fully automating away the review process~\cite{jiang_supporting_2021, fueston_putting_2021}. 

In this work, we instead focus on building tools that can better support the user-driven process of deeply understanding collections of research papers for literature review.
Our work complements existing efforts by enabling researchers to iteratively build and refine structured representations that capture meaningful connections across large paper collections.
\section{Designing LLM-Assisted Workflows for Literature Review}\label{sec:designGoals}
Our design goals are motivated by gaps in prior work in user-driven literature understanding tools detailed in Section~\ref{sec:small_set} and \ref{sec:large_set}. We also draw from cognitive psychology theories on how knowledge workers make sense of large amounts of information, specifically, sensemaking and information foraging theory.

Literature review requires identifying, understanding, and organizing multi-faceted relationships across many research papers~\cite{snyder2019literature, hashimoto2017automatic}. We are inspired primarily by~\citet{pirolli2005sensemaking}'s notional model of sensemaking, which describes how analysts iteratively collect, organize, and synthesize information to generate insights. This framework conceptualizes effective information analysis as ``the process of creating a representation or \textit{schema} to answer task-specific questions,'' where schemas are structured, often externalized, knowledge artifacts created and refined during sensemaking~\cite{russell_cost_1993}. Applied to literature review, such structured representations can scaffold the cognitive progression~\cite{anderson2001bloomrevised} from low-level information retrieval to higher-order processes of analyzing, evaluating, and synthesizing in scholarly sensemaking.

Information foraging theory~\cite{pirolli1999information} further highlights two forms of \textit{environmental enrichment} that can optimize gathering relevant information from information patches (e.g., research papers)---1) reduce the cost of navigating between patches, and 2) make patches yield better returns of valuable information. A structured literature review table serves both these purposes: it functions as an externalized schema for sensemaking while organizing information across papers into a unified view (reducing navigation costs) and extracting key findings (improving information yield). These theories guide our following three design goals for LLM-assisted literature review workflows, addressing common challenges in literature review processes observed in prior work~\cite{khalil2022challenges, daniel2022common, chen2016challenges, thomas2017livingsystematicreviews2, fok2025livingNarrativeReviews}.

\subsection{Design Goals}

\subsubsection*{\textbf{DG1. Help researchers transform a large, unstructured literature collection into a structured repository of extracted relevant information.}}

Literature review requires researchers to process large volumes of unstructured content into meaningful, structured representations. A system should reduce the cognitive costs of extracting, organizing, and accessing relevant information across multiple papers. By creating a structured repository, i.e., a literature review table, that facilitates organization and easy navigation of key information across papers, researchers can shift their mental resources toward higher-level processes of evaluation and synthesis, rather than the tedious and repetitive process of information extraction.

\subsubsection*{\textbf{DG2. Help researchers better make sense of vast information across many papers, transitioning between low-level details and high-level patterns.}} Analyzing patterns and deriving insight across multiple papers is one of the key cognitive challenges in effective literature review, particularly as the number of papers in a review grows. A system should aim to support researchers in making sense of their information repository, and particularly in navigating between successive levels of analysis and structure, e.g., from individual paper review to facet-grounded comparisons to collection-wide synthesis. This could involve providing additional structures that transform the literature review table to scaffold and guide researcher judgment of extracted information, or surfacing potential patterns across papers while enabling researchers to examine and evaluate these suggestions based on their expertise.

\subsubsection*{\textbf{DG3. Help researchers quickly validate the veracity of AI-generated content.}}
Oversight and evaluation of information processed by an AI system and not researchers themselves is critical in maintaining the integrity of a literature review.
As LLMs become increasingly integrate into these workflows---suggesting, extracting, and summarizing information---appropriate mechanisms are needed for researchers to efficiently verify the assistance these models provide.

\begin{figure*}[ht]
    \centering
    \includegraphics[width=0.95\textwidth]{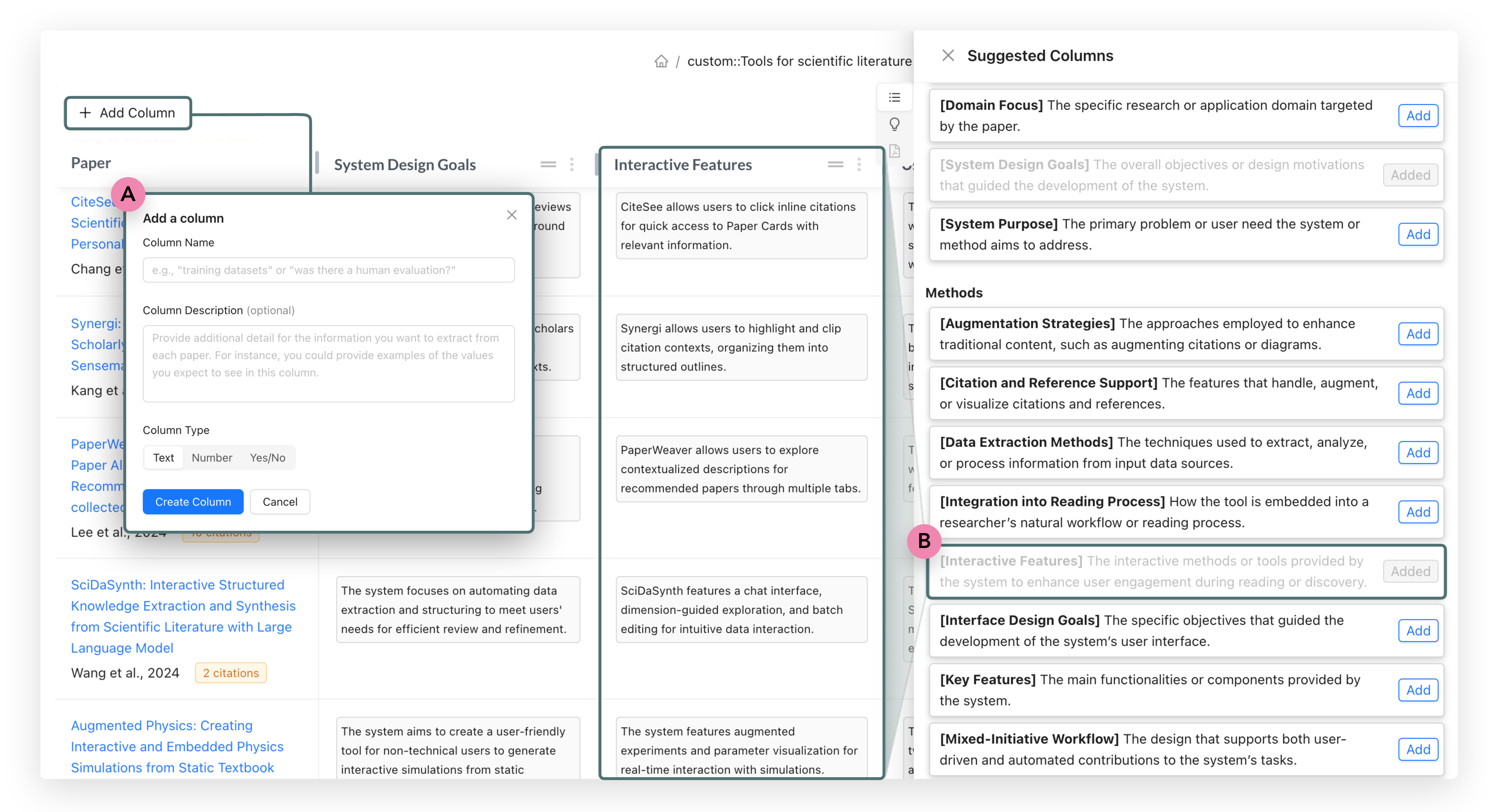}
    \caption{Columns can be added to the literature review table in two ways: A) \textit{User-defined} columns precisely specify a faceted information need and allow additional context for steering LLM assistance; B) \textit{System-suggested} columns offer collection-aware recommendations for columns that can be added with a single click.}
    \Description{}
    \label{fig:createColumn}
\end{figure*}
\begin{figure*}[ht]
    \centering
    \includegraphics[width=0.95\textwidth]{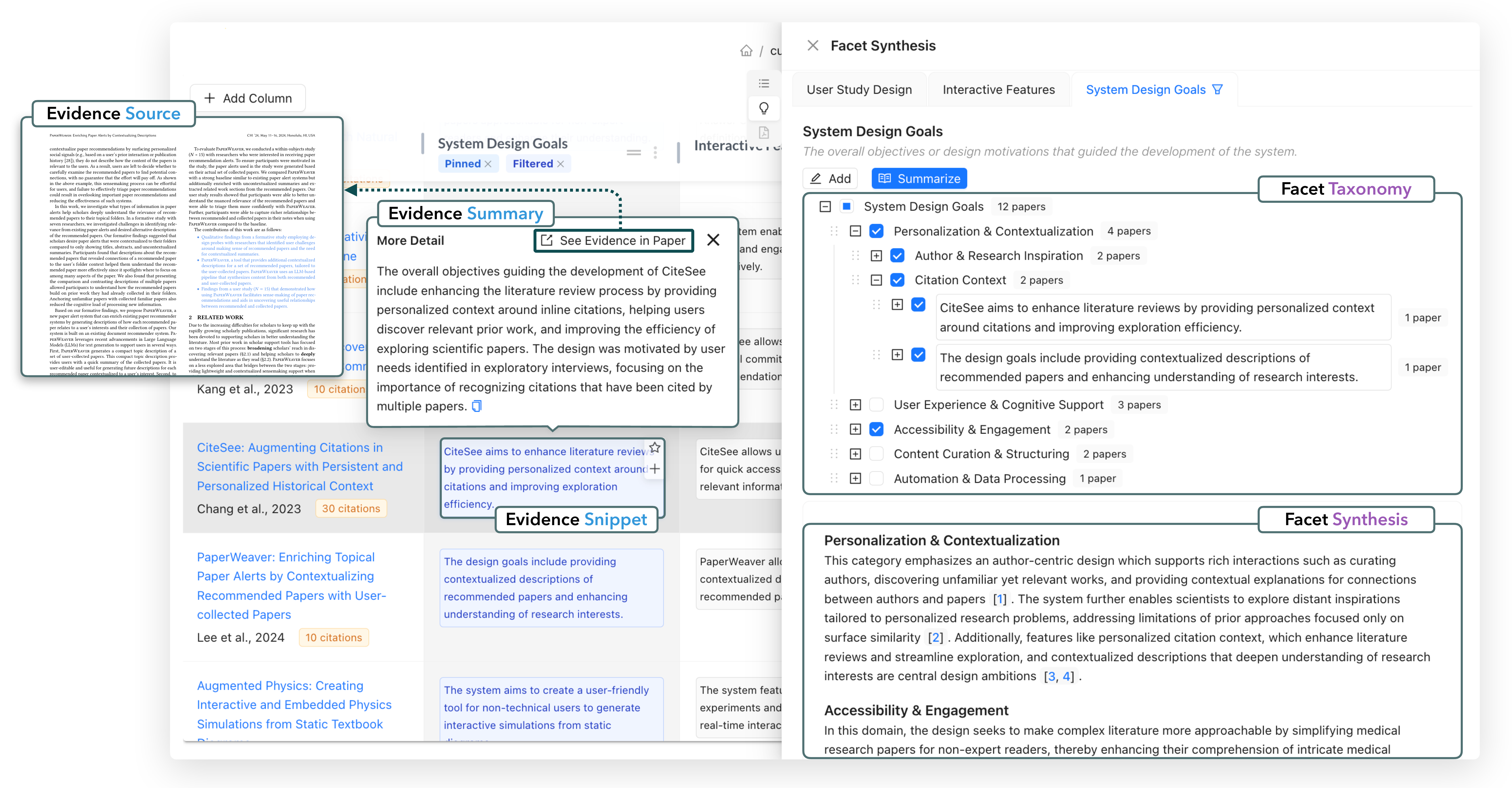}
    \caption{In \sys{}, users review large paper collections by navigating and analyzing information across various structured representations. Each cell in the literature review table is a snippet of faceted information from a paper (evidence snippet). Clicking on a snippet shows a popover with additional detail (evidence summary), with a button that can further open the paper PDF in an integrated paper reader with attributed paragraphs highlighted (evidence source). Faceted columns are transformed into distinct hierarchical taxonomies (facet taxonomy), which can be explored, refined, and used to controllably generate a narrative summary with citations (facet synthesis).}
    \Description{}
    \label{fig:userInterface}
\end{figure*}

\setlength{\fboxsep}{2pt}

\section{\sys{}}
Based on these design goals, we present \sys{}, an interactive system that supports researchers in exploring and making sense of large literature collections by using large language models (LLMs) to transform dense paper information into a series of linked, structured information representations. With \sys{}, users employ natural language to convey specific information needs, and in response, the system populates a literature review table by creating a column in the table generated using evidence retrieved from the full texts of papers in the collection. \sys{} also suggests several collection-aware columns to help users get started (\textbf{DG1}). \sys{} further organizes the information within each column into a hierarchical taxonomy overviewing the available themes across the collection, and allows controllable generation of a narrative synthesis that bridges the schematization and presentation phases of sensemaking (\textbf{DG2}). Connecting these information abstraction layers through interaction, \sys{} facilitates bi-directional exploration between the raw information within individual papers and broader themes that span across multiple papers (\textbf{DG3}).

\subsection{Interface and Example User Scenario}
To illustrate the design and features of \sys{}, we present a user scenario featuring Juno, a researcher in human-AI interaction who is interested in exploring the use of LLMs as evaluators for complex tasks. Before setting a specific research direction, she decides to consult the literature to better understand the existing work. She starts by compiling relevant papers she had saved and results from an academic search engine using the query \textit{LLM-as-a-Judge}. We join Juno as she uploads her collection of 83 papers to \sys{}.

\subsubsection{Defining faceted columns to extract relevant information across papers.}
To start, \sys{} transforms her paper collection into a \textbf{literature review table} with a single \textit{Paper} column, where each cell contains relevant metadata for a single paper (title, author, and citation count). Based on her prior knowledge and reading the papers' titles, Juno has several questions she's interested in exploring. For example, she had noticed two papers in her collection that use LLMs as evaluators for scientific ideation, and wonders, ``\textit{Where else has an LLM-as-a-Judge paradigm been used?}'' She clicks \textcolor{gray}{\fbox{\textcolor{black}{Add Column}}} to create a \textit{user-defined column} in the table representing the information facet she wants to explore (Figure~\ref{fig:createColumn}). In the column creation modal, she specifies the facet (\textit{Application Area}) in the Column Name, offers specific examples (``\textit{scientific ideation, creativity, etc}'') in the Column Description to guide the information generated by the system, and leaves the Column Type as the default type of text.

Using Juno's specification, the system adds a new column to the table and populates each cell in the column with a short snippet of relevant information (\textbf{evidence snippet}) generated from the full text of the corresponding row's paper. Juno scrolls through the table, scanning keywords within the new column. She notices several commonalities at a glance---multiple papers have used LLM-as-a-Judge in applications across science, finance, and law.

\subsubsection{Exploring system-suggested columns}
Next, Juno checks out \sys{}'s \textit{system-suggested columns} by opening a side panel anchored to the right of the screen, containing a list of up to 20 suggested facets that are tailored to her specific paper collection (Figure~\ref{fig:createColumn}). These suggestions could serve to guide novice researchers in exploring an unfamiliar collection, while also aiding experts who may find it easier to recognize facets of interest than to recall them from memory. Each suggested facet has a short name in bold and an italicized sentence-long description. While browsing the list, Juno notices several relevant facets she hadn't considered. She clicks the \textcolor{gray}{\fbox{\textcolor{black}{Add}}} button for one labeled \textit{Challenges and Limitations}, prompting the system to add the faceted column to her literature review table and start extracting relevant information.

\subsubsection{Information scent and progressively disclosed details}
As she scans over the new column, one snippet catches her attention: ``\textit{Like human evaluators, LLMs evaluations are also found to have certain biases}.'' \sys{} shows concise snippets in each table cell by default to reduce information overload, opting instead to progressively disclose relevant details on demand. Juno clicks on the snippet in the table, showing a popover with a paragraph of additional detail (\textbf{evidence summary}), which after reading, she finds has satisfied her previous information curiosity. If she wants to read more about these biases in the authors' own words, she can click the \textcolor{gray}{\fbox{\textcolor{black}{See in Paper}}} button to open the paper's PDF in an integrated paper reader within the side panel. When opened this way, the reader also highlights a block of text relevant to the snippet, guiding her attention and offering a useful entry point into the full text for a deeper dive (\textbf{evidence source}). Instead, she decides to click on the title in the \textit{Paper} column, revealing a popover with the paper's abstract. She skims over the abstract, making a mental note for her future self of how the paper offers a unique angle on biases in LLM-based evaluations. Together, these snippets in the table offer information scent, while the various levels of detail revealed through interaction allow Juno to control her own depth and direction of exploration.

\begin{figure}[ht]
    \centering
    \includegraphics[width=0.48\textwidth]{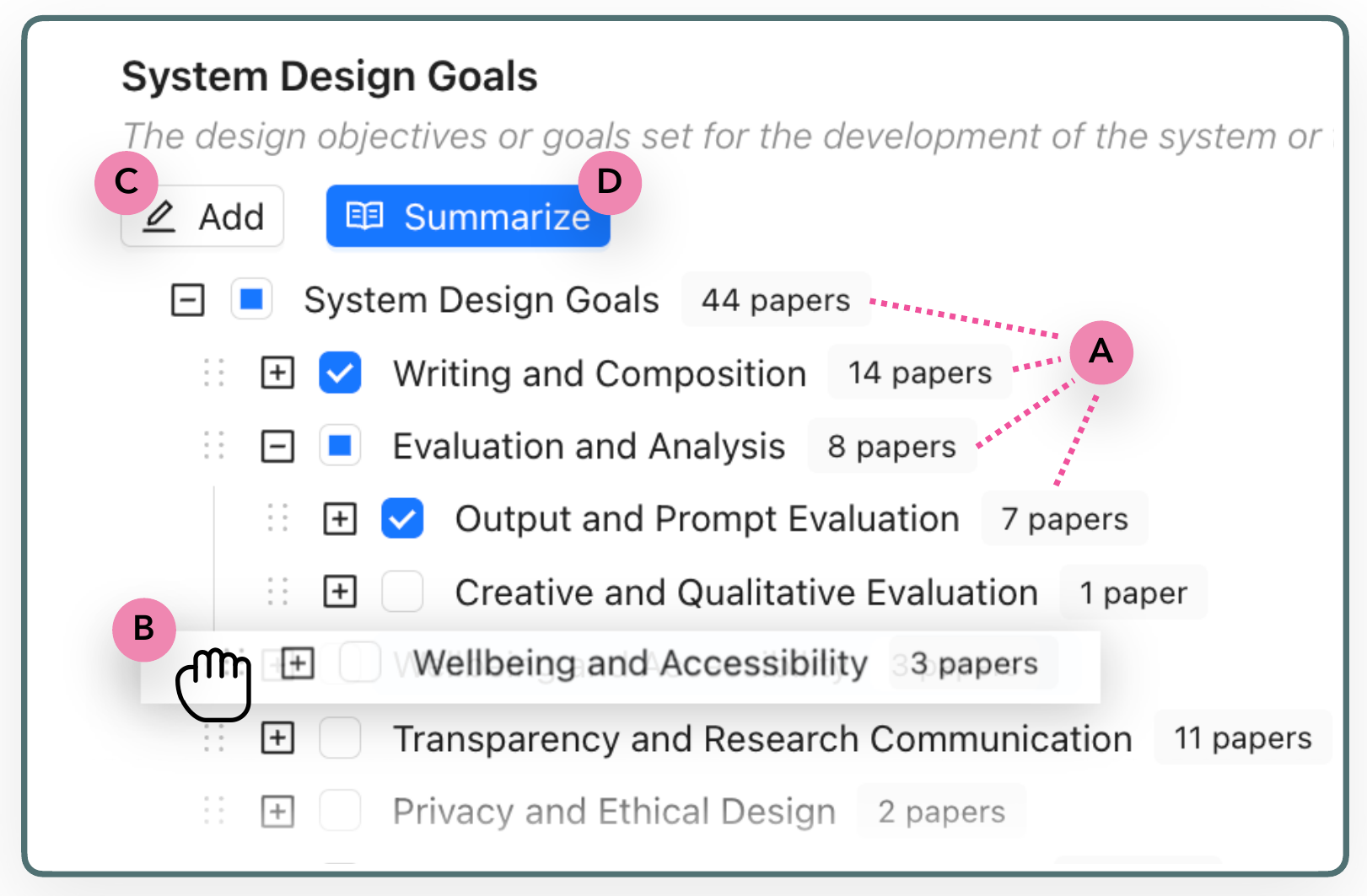}
    \caption{The facet taxonomy. Each category shows the number of included papers (A). Users can manually refine the taxonomy through drag-and-drop interactions (B) or add additional categories (C). If at least one category is selected, the taxonomy can be summarized into prose (D).}
    \Description{}
    \label{fig:taxonomyInteractions}
\end{figure}
\begin{figure*}[ht]
    \centering
    \includegraphics[width=1\textwidth]{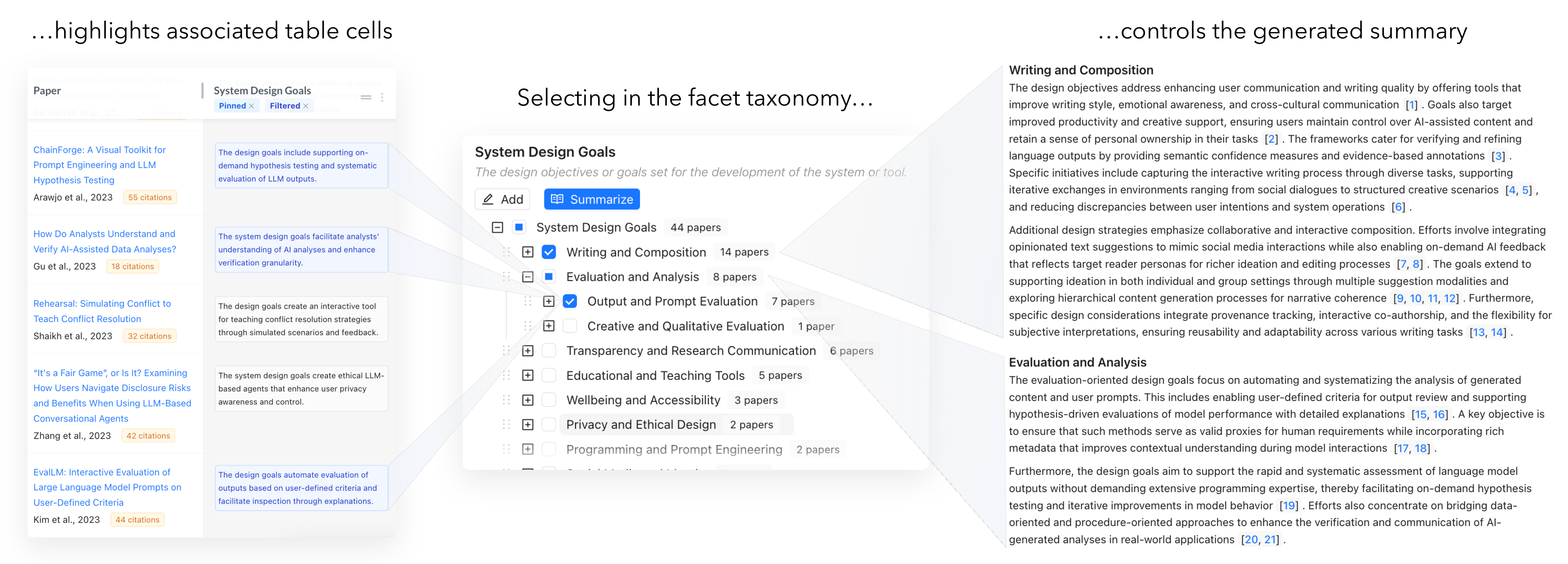}
    \caption{Selecting specific categories in the facet taxonomy: 1) highlights cells for the included papers in the literature review table, allowing users to quickly delineate between and browse the selected (and not selected) papers; 2) controls the structure and papers included in the generated summary.}
    \Description{}
    \label{fig:selectTaxonomyNodes}
\end{figure*}
\begin{figure*}[ht]
    \centering
    \includegraphics[width=1\textwidth]{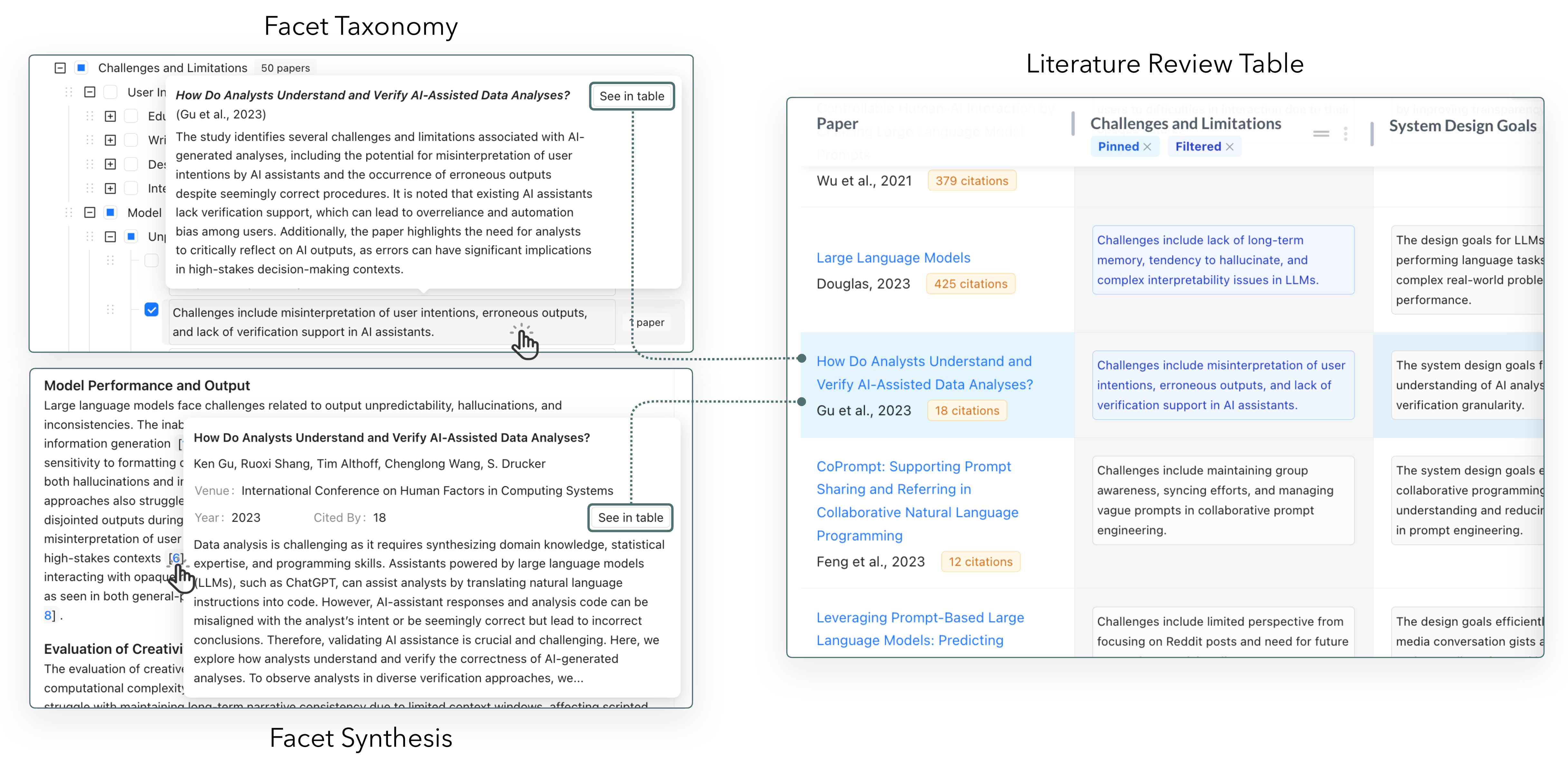}
    \caption{Users can view additional detail while exploring the synthesized representations: 1) Clicking an evidence snippet in the facet taxonomy shows the full evidence summary; 2) Clicking a citation in the facet synthesis shows an in-situ citation card. From either, users can click \textit{See in Table} to scroll directly to the corresponding row in the literature review table.}
    \Description{}
    \label{fig:linkToTable}
\end{figure*}

\subsubsection{Facet-focused sensemaking across documents}
The facet columns allowed Juno to quickly extract relevant information across many papers and served as information scent to drill-down and read specific parts of each paper. However, it can still be difficult to go through the extracted values across all 83 papers to identify higher level themes and their distribution. For this, as Juno adds columns to the table, the system automatically organizes extracted information within each facet into a hierarchical \textbf{facet taxonomy} (one per column) (Figure~\ref{fig:taxonomyInteractions}). Switching to the synthesis panel, she selects a tab for \textit{Application Area}, the first column she added, revealing a tree-like taxonomy grouping related snippets under high-level categories. Each category in the taxonomy displays the number of associated papers, and the taxonomy is sorted to place common categories with more papers at the top, allowing her a quick overview of the information landscape.

Juno begins to explore the taxonomy, scanning the overarching categories before expanding specific ones to reveal their subcategories. Selecting a category highlights the corresponding snippets in the table (Figure~\ref{fig:selectTaxonomyNodes}), allowing her to quickly identify which papers are included in each category, ground the category in lower-level representations in the table (i.e., information extracted from each paper), and potentially reason over any gaps in coverage. While most categories appear appropriate, Juno notices two sub-categories in different levels that should be grouped together. She drags one category to the other, merging them under the same parent. As she continues refining the taxonomy, she shapes both the system’s organization and her own understanding along this facet.

Now satisfied with the structure, she selects the top three categories, comprising the majority of papers in her collection, and clicks the \textcolor{gray}{\fbox{\textcolor{black}{Summarize}}} button, prompting the system to generate a \textbf{facet synthesis} aligned with her refined taxonomy (Figure~\ref{fig:selectTaxonomyNodes}). The synthesis presents a summary generated using only the papers from the selected taxonomy categories following the taxonomy structure, with inline citations which can be clicked to reveal citation cards containing the paper's metadata and abstract, allowing a quick, in-situ assessment of relevance (Figure~\ref{fig:linkToTable}). She saves her work by copying the generated summary with references in a single click, exporting it to her note-taking document for future use.

\subsection{Technical Implementation}
In this section, we describe \sys{}'s computational pipeline, comprising four LLM-enabled components: \textbf{Facet Discovery} (\S\ref{sec:pipeline_facetDiscovery}) in which collection-aware comparative dimensions are generated, \textbf{Value Extraction} (\S\ref{sec:pipeline_valueExtraction}) in which faceted information is retrieved from and attributed to papers' full texts, \textbf{Taxonomy Creation} (\S\ref{sec:pipeline_taxonomyCreation}) in which information within a facet is clustered into emergent themes, and \textbf{Synthesis} (\S\ref{sec:pipeline_synthesis}) in which information is organized into a coherent narrative for presentation.

\subsubsection{Facet Discovery} \label{sec:pipeline_facetDiscovery}
\sys{} automatically identifies high-level comparative dimensions (facets) that serve as analytical lenses through which users can jumpstart or deepen their exploration of the literature. To induce these facets, \sys{} employs a three-stage process grounded in the paper collection. First, we randomly sample $n$ subsets, each containing $k$ papers from the overall collection. For each subset, we prompt an LLM to generate candidate facets using context formed from combining the titles and abstracts of papers in the subset. Finally, we use an LLM to consolidate these $n$ sets of candidate facets into a cohesive final set, prioritizing facets that appear across multiple subsets and merging semantically similar ones. We empirically chose $k=4$ papers per subset---we found larger subsets led the LLM to produce many generic facets in an effort to form dense connections (e.g., ``Main contribution'' or ``Findings'') that, while informative, lacked the desired specificity for analysis. Similarly, we found $n=4$ subsets sufficient, as more subsets rarely yielded additional unique facets.

\subsubsection{Value Extraction} \label{sec:pipeline_valueExtraction}
When a user defines a facet of information they wish to explore (e.g., by creating a new column in the table), we use a retrieval-augmented LLM to generate a paragraph of relevant information from each paper's full text. To avoid overwhelming users as the literature review table grows, we use a second LLM call to distill these dense paragraphs into single sentences. These LLM calls are heavily parallelized to minimize the time users must wait for values to be populated in the table after specifying a column. When users click to view evidence for a snippet attributed to the source PDF, the system encodes both the snippet and full text chunks using a text embedding model (\texttt{all-MiniLM-L6-v2}) and then highlights the chunk with maximal cosine similarity.

\subsubsection{Taxonomy Creation} \label{sec:pipeline_taxonomyCreation}
To connect information across papers within a facet, we use an LLM to cluster the generated snippets into a hierarchical taxonomy. This taxonomy creates a nested tree structure with a dynamic depth, adapting to the natural organization of the information (up to a maximum of 5 levels). Each level represents increasingly specific categorizations, with leaf nodes containing the actual information snippets. The LLM is directed to create meaningful categories that avoid broad categories or excessive fragmentation, such that each category aims to have a reasonable number of conceptually similar snippets. Every paper is required to be included in the taxonomy, and information from a single paper may span multiple categories when appropriate.

\subsubsection{Synthesis} \label{sec:pipeline_synthesis}
The final component supports transformation of the facet-specific hierarchical taxonomy into a coherent narrative, making it more accessible for reuse and sharing. We use an LLM to generate a structured summary that follows the organization of selected branches in the taxonomy, allowing users to steer the narrative based on their specific research interests. The LLM is instructed to ensure their synthesis includes all papers associated with the selected nodes and that any generated claims are explicitly attributed, providing citations immediately after each claim. The result is a comprehensive and tightly attributed summary that aggregates information across papers while clearly tracing information back to its source.

\subsubsection{Additional Details}
\sys{} is implemented as a web application with Flask (Python) backend and React (Typescript) frontend. Paper collections can be created in the system from a list of titles or Semantic Scholar IDs, or created interactively from a search query, with relevant papers fetched using Semantic Scholar's paper relevance search\footnote{\url{https://api.semanticscholar.org/api-docs/graph\#tag/Paper-Data/operation/get_graph_paper_relevance_search}}. Given a collection, the system downloads all available open-access PDFs and uses GROBID~\cite{GROBID} to parse full text, section data, and token bounding boxes from the PDFs. Semantic Scholar APIs are used for fetching paper metadata, including authors, year, venue, and citation count. Most LLM-enabled components use OpenAI's \texttt{o3-mini} (with ``low'' reasoning), except for value extraction which uses \texttt{GPT-4o-mini}. The specific models and parameters were selected to balance usable interaction latency and performance for each component. Additional details and LLM prompts are available in Appendix~\ref{appendix:prompts}.

\section{User Study}
\begin{table*}[t]
    \centering
    \small
    \begin{tabular}{lll}
        \toprule
        \textbf{Survey Paper} & \textbf{Topic} & \textbf{Taxonomy Dimensions} \\
        \midrule
        Lee et al.~\cite{lee2024writingassistants} & Intelligent and interactive writing assistants & \textbf{Task}, User, Interaction, \textbf{Technology}, Ecosystem \\
        Pang et al.~\cite{pang2025llmification} & Use of LLMs in HCI research & \textbf{Application Domains}, LLM Roles, \textbf{Limitations \& Risks} \\
        \bottomrule
    \end{tabular}
    \caption{The survey topics and taxonomy dimensions used in the tasks, and their source papers.}
    \label{tab:taskPapers}
\end{table*}

We conducted a within-subjects user study with 23 researchers to evaluate the effectiveness and usability of \sys{}, in contrast to a more conventional workflow involving manual review with conversational LLM assistance. Our evaluation aimed to answer the following questions:
\begin{enumerate}[label=\textbf{RQ\arabic*.}]
    \item How does \sys{}'s workflow of scaffolding literature review with generated structured representations compare to a manual approach assisted by an LLM chat-based baseline?
    \item How do researchers leverage different structured representations provided by \sys{} to make sense of information scattered across large collections of research papers?
\end{enumerate}

\subsection{Participants}
We recruited 23 computer science researchers (13 female, 10 male; Age: $M=27$, $SD=4$) via university mailing lists, social media recruitment messages, and snowball sampling. Based on a screening survey, we filtered out those who self-reported no prior research experience or were not at all comfortable with reading abstracts and papers in HCI, since our study involved literature review tasks within this domain. Otherwise, participants were recruited on a first-come, first-served basis. Participants consisted primarily of PhD students (17/23), with most participants having at least 3 to 5 years of academic research experience (17/23). Participants' primary research areas were largely HCI (18/23), with focuses in AR/VR, accessibility, human-AI interaction, and AI ethics, among others. All participants were based in the United States. Most participants (19/23) used LLM-based applications at least weekly for general tasks, with 12 using them daily. Usage of LLMs for research varied, with 14 reporting extensive use of LLMs for research, 6 who used LLMs occasionally, and three who used them rarely or never. Additional participant details are available in Appendix~\ref{appendix:participantDemographics}.

\subsection{Task}
We designed our study tasks to closely resemble the initial process of drafting an outline for organizing related literature, e.g., when writing a literature review or survey paper. Given our within-subjects study design, each participant used both system conditions, but with a different task, as repeating the same task would introduce learning effects. The two tasks were designed to be similar in difficulty and nature, allowing for valid comparisons of system performance across different tasks. For each task, we selected a survey paper in HCI~\cite{lee2024writingassistants, pang2025llmification}, extracted the top-level taxonomy explicitly defined in the paper, and sampled 50 papers from the surveyed literature. The collection size of 50 papers ensured consistency across tasks while balancing sufficient material for meaningful organization within the study timeframe. The selected taxonomy dimensions similarly balanced accessibility and depth across the two tasks: one dimension could be reasonably developed from abstracts (Task/Application Domains), while the other required deeper engagement with the full texts (Technology/Limitations \& Risks). Participants were given the paper collection and survey topic, then asked to continue filling in the outline for the assigned sections (bolded). Their goal was to create meaningful subsections, save relevant information, and cite appropriate papers. Additional details are provided in Appendix~\ref{appendix:taskDetails}.

\subsection{Baseline}
In the baseline condition (\textsc{Baseline}), participants used a system that displayed a searchable list of paper metadata---including title, authors, year, venue, abstract, and a PDF link. This design emulated the standard experience of browsing conference proceedings or results from an academic search engine. To reflect the growing integration of LLM-based tools in scholarly workflows~\cite{liao2024llmsresearchtoolslarge}, participants were encouraged to use ChatGPT,\footnote{\url{https://chatgpt.com/}} a commercial LLM chat application, to support their exploration.\footnote{During the study, participants were provided with login credentials for a ChatGPT account created exclusively for this user study. This served two purposes: to standardize the version of ChatGPT each participant used and to allow us to export detailed logs of participants' prompts and LLM responses.} They could interact with the model via text prompts but were restricted from web search (since finding additional papers was beyond the scope of the task).

\subsection{Procedure}
The first author (study facilitator) conducted studies remotely with participants over Google Meet in March 2025. Each study lasted around 90 minutes. The study began with an introduction and overview of the two systems. Participants were first given a brief tutorial of \sys{}, and then allowed to use it to explore a sample paper collection involving ``\textit{interactive tools for scientific literature review}.'' This familiarization period lasted up to ten minutes, during which the study facilitator answered any questions. Then, participants were asked to explore the same collection in the baseline interface. No tutorial was given for ChatGPT since all participants indicated in their screening survey some prior familiarity with LLM-based chat applications. Next, participants worked on the first task with the assigned system for 30 minutes, after which they completed a post-task survey. This process was repeated with the second task and assigned system. The pairing of task and system was fully counterbalanced across participants. After completing both tasks, any remaining time was used as a semi-structured interview in which participants elaborated on and compared their experience using both systems. The study facilitator also probed into any interesting usage behaviors participants exhibited during the study. Participants received a \$60 USD gift card upon completion of the study. This study was reviewed and exempted by our organization's internal review board.

\subsection{Measures and Analyses}
For quantitative data, we collected participants' post-task survey responses, including perceived system effectiveness, user control, information transparency, and overall satisfaction with the literature review process using each system (see Appendix~\ref{appendix:postTaskSurveyQuestions} for the specific questions). We also captured and analyzed interaction logs from the sessions to measure the quantity and types of actions participants took (e.g., create a user-defined column, add a system-suggested column, select or refine taxonomy categories, summarize). We did not analyze the content of the outlines participants created during their tasks due to the significant diversity in form and content, complicating an unbiased expert evaluation. Instead, we treated the outline creation as a contextual activity that helped participants engage with the task while using each system, and relied on self-reporting and quantitative measures in the behavioral logs to examine the effectiveness of the two systems. Finally, participants' exit interviews were recorded and automatically transcribed. To analyze all paired Likert-scale data, we conducted Wilcoxon signed-rank tests with Bonferroni corrections for multiple comparisons. For qualitative data, we conducted a reflexive thematic analysis~\cite{braun_thematicAnalysis_2006}, where the first author developed and iteratively refined codes through discussions with the research team to identify emerging themes.

\section{Results}

\begin{figure*}[t]
    \centering
    \includegraphics[width=0.95\textwidth]{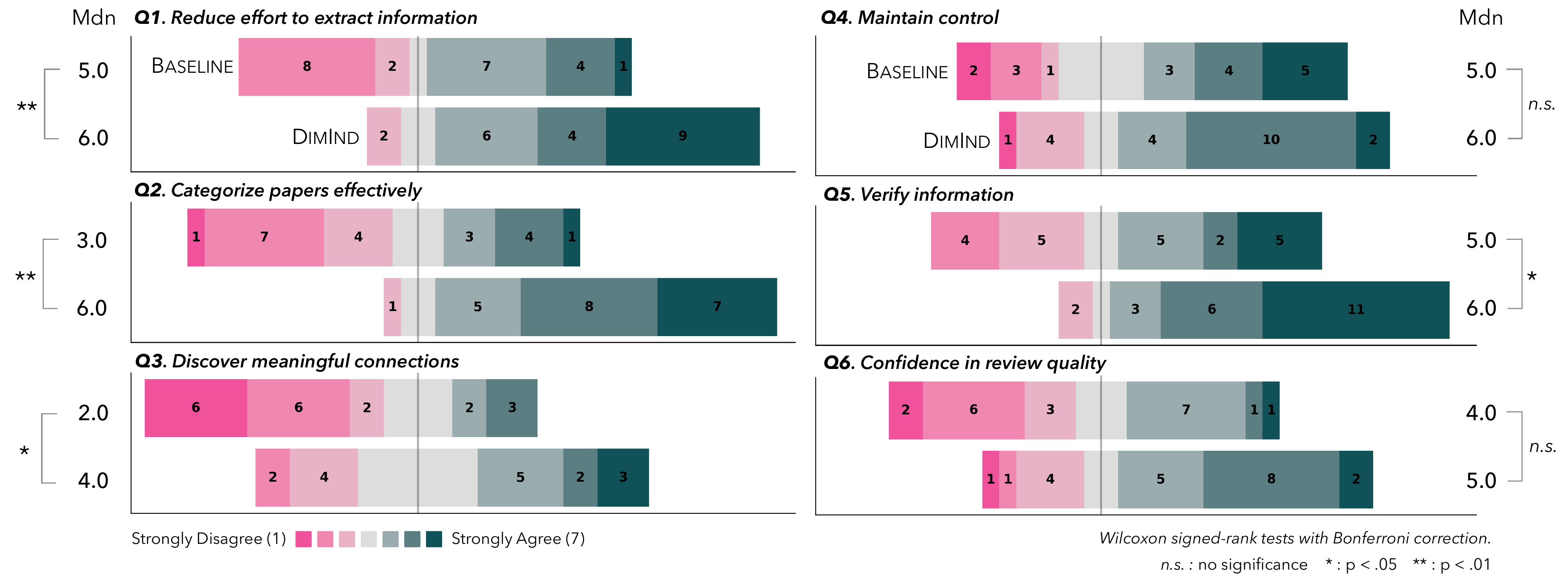}
    \caption{Participants' post-task ratings (7-point Likert scale) of literature review utility, control, information verifiability, and confidence across \sys{} and \textsc{Baseline}.
    }
    \Description{}
    \label{fig:surveyResults}
\end{figure*}

In this section, we present our quantitative and qualitative findings, organized by our two research questions. Participants are referred to with pseudonyms P1--P23. Quotes were lightly edited for brevity and clarity. We denote the median rating of \textsc{Baseline} and \sys{} as $M_{B}$ and $M_{D}$, respectively, and the Wilcoxon test statistic as \textit{W}. Statistical significance was determined at $p < .05$.

\subsection{How effectively does \sys{} support deeply exploring large collections of papers? (RQ1)}
\subsubsection{Tables with extracted information effectively alleviate foraging costs but risk information overload}
Based on the post-task ratings, we found participants appreciated the ability to quickly extract information from many papers at once through the creation of custom and system-suggested columns in \sys{}, reporting less perceived mental effort in extracting and organizing relevant information ($W=21.5$, $p<.01^{**}$, $r=0.74$) with \sys{} ($M_D=6$, $IQR=5-7$) than with \textsc{Baseline} ($M_B=5$, $IQR=2-5$) (Fig.~\ref{fig:surveyResults} -- Q1). While many participants were able to eventually extract similar information with \textsc{Baseline} with much more effort, they appreciated \sys{}'s support for easily and explicitly creating custom columns that matched the specific questions they had, and to have all the information organized together within the familiar table representation (P2, P13, P14, P18, P21). Contrastly, without careful prompting the baseline system often generated lots of information to mixed utility. More specifically, when using the baseline system, some participants felt they had to sift through information that ''\textit{tended to be a bit overly fluffy}'' (P15), even to the point of irritation as P19 expressed: ``\textit{These lines are just kind of making me mad\dots`AI can act like the third teammate', it doesn't really tell me anything.}''

While participants appreciated being able to quickly extract relevant information across large collections of documents efficiently, we also observed a different flavor of information overload in \sys{}. Even when the information presented by the system was seen more useful, presenting it all at once without the users having to do much work can be overwhelming. P19 expressed ambivalent feelings, finding the table ``\textit{very useful to cut down the manual labor}'' but also how ``\textit{having everything ready for me all at once can be overwhelming, since whenever I'm doing this it's usually one row at a time.}'' When reviewing large collections, even a relevant snippet for each paper felt unapproachable to use in drawing connections and organizing ideas across papers, as P4 described, ``\textit{it's too detailed for me to organize it just by reading through these cells.}'' Rather than directly interacting with the extracted information, P13 viewed the extracted information more as an auxiliary ``\textit{database to refer back to with all the columns and information}'' when writing their literature review, and only then selectively investigating cells for papers they intended to discuss.

Instead of trying to consume all the information from the system at once, most participants using the LLM-generated information as \textit{information scent}~\cite{pirolli1999information}, where keywords and categorizations served as navigational aids rather than as content directly integrated into a review. As P2 explained, ``\textit{these keywords are good enough for me to quickly skim through the PDF and find where it might lie},'' suggesting that the generated information helped guide attention to relevant evidence within papers. For some, these cues served primarily as navigation---further interpretation would necessitate reading ``\textit{directly from the author's mouth instead of a paraphrased version}'' (P2).

Overall, these ``\textit{jumping-off points}'' (P15) in the extracted evidence snippets facilitated efficient exploration, directing participants to relevant papers while enabling them to browse extracted information for the same facets across multiple papers simultaneously. For example, while P5 was examining a \textit{Research Focus} column they had added to the table, they came across one snippet that mentioned ``potential and risks of LLMs.'' To see more context, they clicked to open the paper's PDF, and the system scrolled to and highlighted the first paragraph of the discussion. They then briefly continued to browse snippets for other papers in the table, before directing their attention back to reader to skim the discussion section of the paper they had previously opened. 

\subsubsection{Taxonomies as cognitive scaffolds for thematic exploration}
While participants leveraged the table structure as \emph{information scent} for the underlying documents, most looked to the taxonomy structure in \sys{} for supporting higher-level synthesis. Overall, participants rated \sys{} ($M_D=6$, $IQR=5-7$) higher than \textsc{Baseline} ($M_B=5$, $IQR=2-5$) for its ability to support paper categorization ($W=0$, $p<.01^{**}$, $r=0.88$) (Fig.~\ref{fig:surveyResults} -- Q2).\footnote{$W=0$ indicates all participants rated \sys{}'s support for paper categorization greater than or equal to that of \textsc{Baseline}.} Many participants attributed this advantage to \sys{}'s taxonomy representation clearly organizing emergent themes and relevant papers into an interactive hierarchical tree. P23 shared how the transformation from faceted column to taxonomy provided an efficient way to summarize themes across many papers for a facet:

\begin{quote}
``\textit{The tree structure was unexpectedly useful as a way to view themes\dots With [\textnormal{\texttt{Baseline}}], you would have to compile relevant abstracts, titles, and metadata yourself before feeding it into the LLM. With the table [referring to \sys{} as a whole], it was a seamless way to direct information into the models and get thematic summaries.}''
\end{quote}

\noindent During exploration, the taxonomy structured representations served as external cognitive scaffolds, allowing participants to maintain multiple parallel organizing schemes at once.
For example, P13, who had created a table facet about \textit{ethical considerations}, and compared its cell values that included phrases such as \textit{potential misuse of opinionated language models} and \textit{perpetuation of societal biases} and the taxonomy tree that included higher-level themes such as \textit{Bias}, \textit{Transparency}, and \textit{Societal Harms}, explained: ``\textit{I like how the tree structure gave a very intuitive way of grouping things together, and the way each tab[table facet] has a different tree structure. So it's a different way of organizing your thoughts.}'' This externalization of thought across different `tabs' each with a unique taxonomy further serves to reduce the cognitive load of multi-faceted analysis.

However, while the taxonomy offered a rapid thematic overview, participants generally felt both \textsc{Baseline} ($M_B=2$, $IQR=1.5-4$ and \sys{} ($M_D=4$, $IQR=3.5-5$) fell short of helping them discover very specific and meaningful connections (Fig.~\ref{fig:surveyResults} -- Q3), ones they would have been unable to discover without LLM assistance ($W=15.5$, $p<.05^{*}$, $r=0.78$). 
Combined with the length of the study, we found participants exhibited mixed levels of confidence in the quality and comprehensiveness of the outline they created (Fig.~\ref{fig:surveyResults} -- Q6); while participants reported slightly higher confidence in their outline using \sys{} ($M_D=5$, $IQR=3.5-6$) than \textsc{Baseline} ($M_B=4$, $IQR=2-5$), this difference was not significant ($W=34.5$, $p=.08$, $r=0.66$). This is not surprising---while results are promising that \sys{} is able to significantly increase efficiency when deeply exploring a large collection of papers, the task of comprehensive literature reviews would still require significant time and effort beyond the scope of our lab study.

In sum, current models still fall short in generating deep intellectual connections between documents, highlighting the continued importance of a researcher-driven analytical process. While our findings are promising with regards to efficiency, designing for longer-term effects on confidence and the impact of prolonged usage remain directions for future work.
At the same time, while most prior work focused on support literature understanding using a single representation, our results point to benefits of providing different structured representation, and the importance of allowing users to fluidly ``zoom'' between different levels of compression, allowing \sys{} to scale literature understanding support to large collections of papers.

\subsubsection{Desire to blend structured representations and conversation}
While structured representations in \sys{} provided valuable scaffolding and more predictable interaction outcomes for the literature review task, participants also valued the adaptability and flexibility offered by conversational interaction in \textsc{Baseline}.
We found no significant difference ($W=65.5$, $p=1$, $r=0.78$) between \sys{} ($M_D=6$, $IQR=4-6$) and \textsc{Baseline} ($M_B=5$, $IQR=3.5-6$) with respect to the ability to maintain control over the literature review process when aided by LLM assistance (Fig.~\ref{fig:surveyResults} -- Q4).
Participants saw advantages and limitations of both interaction paradigms, and instead envisioned an ideal system that would blend structural organization with conversational control. For instance, participants appreciated the ability to trace information across structured representations, specifically drilling down into the specific evidence within a paper's PDF and referring back to the paper's abstract within the table and synthesis. As a result, participants reported finding it slightly easier ($W=10.5$, $p<.05^{*}$, $r=0.81$) to verify system-generated information when desired using \sys{} ($M_D=6$, $IQR=5.5-7$) than \textsc{Baseline} ($M_B=5$, $IQR=3-6$) (Fig.~\ref{fig:surveyResults} -- Q5). At the same time, they also suggested several ways in which \sys{} could benefit from conversation, for example, to follow up with clarifying questions (P17), request explanations for certain decisions (e.g., why certain categories were created in the taxonomy and not others) (P15), explore relationships between specific papers (P23), or modify output formats and detail levels (P9).

\subsection{How do researchers use and navigate across \sys{}'s structured representations? (RQ2)}

\subsubsection{Balanced usage of user-defined and system-suggested facets for creating literature review table}
To transform the unstructured paper collection into a literature review table, most participants (14/23) took a balanced approach in adding both user-defined and system-suggested columns (6 added user-defined columns only; 3 added system-suggested columns only). Participants particularly valued the ability to customize the faceted columns based on their specific criteria. As P2 explained, ``\textit{I really like that I could customize what the criteria for generating an outline could be in that system. Because otherwise just asking an LLM to prompt---it defines its own criteria which might not be how researchers might approach writing [about] the literature.}'' This flexible extraction from many papers at once helped reduce cognitive load, with P6 noting that after defining columns at a high level, their ``\textit{job is reduced down to only checking whether the information is correct.}'' These results further showed the importance of supporting user-driven exploration, and that LLM generated structures might not always cover the nuanced and personal information needs of different users.

\subsubsection{Facet taxonomies as central hubs for exploration}
Facet taxonomies emerged as central navigational hubs during exploration. Participants saw them as \sys{}'s most useful information representation. The majority of participants primarily interacted with the taxonomy during their tasks, expanding and reviewing generated categories and used the taxonomy as \emph{thematic filters} for papers in the table.

Most participants (19/23) used the taxonomy at least once, with participants expanding and collapsing categories an average of 18.0 times (SD=17.1). Similarly, most participants (20/23) selected categories in the taxonomy, using it to focus their attention on specific paper subsets, an average of 16.3 times (SD=18.1). As P4 noted, the tree structure helped overcome the challenge that specific paper evidence in the table representation could be ``\textit{too detailed for me to organize just by reading through},'' adding that ``\textit{after filtering through [the taxonomy] then it made more sense for me to look at what appeared.}'' This behavior shows how participants used taxonomies to transform overwhelming tables into manageable views by first selecting relevant categories before examining filtered paper evidence to understand more deeply each generated category, and to confirm their validity.

Moreover, P23 described how \sys{}'s integrated workflow from specific faceted evidence to taxonomy addressed a challenge of traditional literature review fragmentation where researchers ``\textit{end up with 10 different documents of scattered information and a really clunky spreadsheet.}'' She frequently navigated between taxonomy categories and specific papers by clicking connections between evidence in the taxonomy and table rows, summarizing:
\begin{quote}
\textit{``You're able to scan through a lot of information, you're able to expand things that you want to look at, and then most of the time it's in this very compact form. I found myself using the tree the most, and I really like being able to switch back and forth between looking at the actual text, so the abstract or the whole paper, and then looking at the automatically generated information or summaries.''}
\end{quote}
Altogether, these findings suggest that the faceted taxonomies bring valuable high-level structure to literature review, effectively bridging overview and detailed exploration.

\subsubsection{Transforming facet taxonomies to narrative summaries}
About half of the participants transformed at least one taxonomy into a narrative summary using \sys{} (12/23). Five of these participants interacted with the generated summaries, closely reading and inspecting more detail by clicking the inline citations (clicking these reveal a popover with metadata and the abstract for the cited paper) (P3, P6, P7, P9, P17). The other group of participants generated the summary to use as a reasonable artifact they could easily export into their task document.

Seven participants also tried to further improve the LLM-generated taxonomy, using those revisions to inform a more desirable summary. These participants iteratively refined the categories with the drag-and-drop interaction, often to group categories that the system had split but participants believed made more sense to be analyzed together. Some participants similarly expressed a desire to directly manipulate the taxonomy, but refrained from doing so given their time constraints. LLM-generated categories were not perfect, nor were they expected to be---when P20 noticed a misalignment between the LLM-generated taxonomy and her own understanding, e.g, ``creative writing'' nested under ``creative'' rather than ``writing'', she described a simple repair:
\begin{quote}
``\textit{Sometimes it did classifications that didn't make a ton of sense. With the writing one, where it separated creative writing from the rest of writing for some reason. But that felt a lot more natural. I can see that and be okay, I think of creative writing as writing. So in my own notes, let me just classify it like that.}''
\end{quote}
These findings suggest the utility of LLM-generated categories as an initial guide, with subsequent interactive, in-situ refinement or post-hoc refinement when exporting her understanding for presentation.

Despite engaging with the faceted table and taxonomies, participants were hesitant to delegate the final narrative synthesis to \sys{} (or other LLM tools) for their own work. This hesitation centered around a desire to preserve researcher agency, particularly in transitioning from the organizational schema of categories and relevant papers to a presentable artifact. In the notional model of sensemaking~\cite{pirolli2005sensemaking}, this suggests that most participants spent their time, and preferred LLM assistance, in the information extraction and schema refinement stages and less in the presentation stage. For instance, P20 explicitly preferred maintaining control over the final synthesis, saying ``\textit{I would probably not use the summary tool... That feels a little bit too close to writing my paper for me,}'' while recognizing the utility of \sys{} in earlier stages.

Participants also shared concerns about automating what they saw as an essential scholarly process for intellectual development: ``\textit{Literature review is to figure out what are all the keywords that you don't know but convey the same meaning... having the automated process might accidentally limit some of the literature that you should know},'' and how over-automation with LLMs could ``\textit{kill the serendipity findings}'' that emerge from personal exploration (P19). This suggests a nuanced relationship where researchers value LLM assistance for exploration and organization but still desire preservation of human authorship---a balance between LLM enhancement and maintaining scholarly agency and integrity.

\section{Discussion}
Literature review and synthesis at scale is one of the most cognitively and labor-intensive activities that is core to the research process. Recent advancements in AI---particularly LLM capabilities for processing and extracting information from long, complex documents---has led to growing research interest in leveraging these new technologies to support literature review.

While some prior work has sought to automate the entire process~\cite{susnjak2025automating,kasanishi2023scireviewgen,altmami2022automatic}, we instead found participants often expressed a desire to retain a scholarly voice and steer the literature review process.
Furthermore, user-driven processes can allow scholarly learning, as one participant noted how traditional approaches involving the manual construction of literature review tables can foster deeper engagement with the literature, helping to explore key ideas and possibly discover interesting connections.
Similarly, some participants described appreciating the serendipity, curiosity, and intellectual development afforded by user-driven exploration in a literature review.

While participants found value in the generated table and taxonomy structures that provided information scent and broad overviews of many papers, they were more skeptical of the formulaic style of LLM-generated writing and highlighted ambiguity around intellectual ownership of the output, saying: ``\textit{Working with ChatGPT---yes, the dots are all connected for me, but it just feels very weird to say that this is mine.}''
As a result, they were hesitant to save and share paragraphs directly generated by LLMs in both conditions, citing the potential for a loss of personal voice and unconscious plagiarism.

Compared to prior user-driven literature understanding systems that support exploring smaller sets of papers by only focusing on providing a single structured representation, such as paper pairs~\cite{lee2024paperweaver, murthy2022accord}, tables~\cite{elicit,wang2024scidasynth,newman2024arxivdigestables} or taxonomies~\cite{hsu2024chime,kang_threddy_2022,kang_synergi_2023}, our results instead suggest that providing multiple structured representations of paper information and allowing users to fluidly move between these different levels of information abstraction and compression can be effective in scaling a user-driven and LLM-assisted literature review process. 

Most commonly, after adding a few columns of faceted information into the table, participants used a top-down review approach, structuring their exploration around the generated taxonomy. And while exploration of a facet often began in the taxonomy by reviewing the available concepts and their distribution in the collection, participants frequently navigated information at multiple levels of detail---for instance, interest in a particular taxonomy category could quickly transition into reading detailed evidence for the specific papers in that category within the table representation---to deepen and verify their understanding. At the same time, participants pointed to a gap between the table and taxonomy structures in \sys{}, where deeper connections between small subsets of papers were lacking. This suggests introducing an additional layer of structured representation that leverages prior work can potentially fill this gap. For example, leveraging PaperWeaver~\cite{lee2024paperweaver} or ACCoRD~\cite{murthy2022accord} to help users discover specific and nuanced connections between closely related pairs of papers in the table before looking at broader themes in the taxonomy structure.

We also found that leveraging LLMs to increase the efficiency of user-driven literature review highlighted a shift from interaction to cognitive costs, revealing several new design challenges. Specifically, given a particular facet of interest, the laborious manual process of reading and extracting relevant information from each individual paper can be greatly reduced with LLMs, yet the amount and rate at which information is presented to the user can become cognitively overwhelming.
In \sys{}, we addressed this by presenting multiple levels of abstraction (tables, taxonomy, and narrative synthesis) as cognitive scaffolds, but the design space remains vast. What abstractions or interactive mechanisms best support sensemaking over large volumes of LLM-generated information without overwhelming or impeding serendipitous exploration. And how well do these designs transfer to other domains of information synthesis (e.g., medicine, clinical, or financial decision making) where AI is increasingly being studied and used~\cite{duede2024oil,liao2024llmsresearchtoolslarge}?

A final challenge involves designing to mitigate risks of LLM hallucination at scale. To this end, we designed \sys{} with information transparency as a first-class goal, providing clear traces between transformations of paper information and interactive access to a paper's abstract or full text to more easily verify generated information. Still, enabling verification at scale requires more accurate attribution techniques and more seamless interaction mechanisms. For example, \sys{} can build on recent advances in fine-grained attribution~\cite{chuang2025selfcite, zhang2024longcite, slobodkin2024attributefirst} to more effectively select and highlight relevant evidence when users drill down into the paper itself, or even proactively flag likely hallucinations for user review~\cite{wadden2022scifactopen, wadden2022multivers}.

\subsection{Limitations}
We note several limitations of our evaluation that shape how our findings should be interpreted. Our within-subjects study was limited to 90-minute sessions to minimize participant fatigue, but literature review is inherently a complex and dynamic process that can take days or months to complete. Similarly, we set the number of papers to be explored in each task to be 50 papers sampled from real-world survey papers. While we believe this is a reasonable number balancing participant effort and scale, survey papers would typically reference more than 100 papers. At the same time, we believe this is a sufficient scale such that our findings may be generalizable to real-world literature review scenarios. We hope to continue to improve \sys{} based on participant feedback and conduct a deployment or longitudinal studies where researchers create and explore their own paper collections over extended periods. Such studies could reveal novel usage patterns and limitations of \sys{}, and provide deeper insights into the long-term effects of LLM-assisted cognitive scaffolding. Our evaluation also consisted primarily of CS researchers with an HCI focus; additional studies involving more diverse academic disciplines are necessary to establish broader generalizability and utility of LLM-assisted structures in literature review workflows.

Additionally, while all participants had prior experience with conversational LLM assistance (e.g., ChatGPT) for both everyday tasks and research purposes, \sys{}'s structured approach to LLM-enabled features presented a learning curve. Combined with familiarity and potential biases with conversational interfaces, these factors could have influenced participants' expectations and perceptions of the reliability of ChatGPT and our system for the assigned tasks. Finally, our system's interactive information traces with detailed paper information are only available when PDFs are openly accessible. Otherwise, \sys{} falls back to paper abstracts, which lack sufficient detail for supporting the system's overall goal of enabling the exploration and analysis of more nuanced paper facets.

\subsection{Future Work}
Based on our findings, we highlight several exciting directions for future work. We plan to release \sys{} as an open platform for researchers to further explore these structured representations and continue to improve the underlying LLM mechanisms for facet discovery, taxonomy creation, and synthesis.

\subsubsection{Enhanced User Controls}
Future work could introduce interaction mechanisms that improve steerability of LLM assistance. For instance, participants expressed interest in capabilities that would allow them to directly influence the system's analysis, such as the ability to highlight specific papers or cells of evidence in the table to steer the generated taxonomy and synthesis (P18), hide papers from the analysis, directly edit or tag system-generated evidence (P11, P13, P18), and merge or split faceted columns (P9). These features could better shape the system's output to align with researchers' personalized needs.

\subsubsection{Multi-Faceted Multi-Document Synthesis}
Analysis support in \sys{} is currently limited to single facets using the taxonomy structure, but participants also expressed interest in analyzing relationships between multiple facets. For example, exploring interactions between table facets about \emph{application domain} and \emph{risks} could help discover risks that are understudied in a particular application domain.
Future work could explore interactive 2D pivot visualizations---similar to \cite{suh2024luminate} but where the two axes are themes----across two table facets to find correlations between facets, provide intelligent suggestions for which facets to analyze together, or introduce new interactions for simultaneously examining multiple facets in the table at once.

\subsubsection{Fine-Grained Literature Discovery}
The structured, faceted information in \sys{}'s literature review table present opportunities for more fine-grained approaches to literature discovery. For instance, users could explore papers that share similar values to those found in a specific table cell, connect to broader conceptual clusters in the taxonomy, or introduce novel values that expand the current facet.

\section{Conclusion}
In this paper, we presented \sys{}, an LLM-enabled system that scaffolds literature review by transforming unstructured paper content into navigable structured representations, with provenance of LLM-generated information to support paper-level verification. Our evaluation with 23 researchers demonstrated that \sys{} provides valuable cognitive support, particularly for extracting faceted information across papers and enabling top-down exploration via conceptual taxonomies. Overall, our findings suggest these structured representations offer an effective way to leverage LLM assistance toward supporting literature review at scale.

\bibliographystyle{ACM-Reference-Format}
\bibliography{main}

\appendix
\section{User Study Details}

\subsection{Demographic Details of User Study Participants} \label{appendix:participantDemographics}
Table~\ref{tab:participantDemographics} lists demographic details for all 23 participants included in our user study, including their research area, years of research experience, and use of LLMs for general and research tasks.

\begin{table*}[t]
\centering
\small
\begin{tabular}{llllllll}
\hline
\textbf{ID} & \textbf{Age} & \textbf{Gender} & \textbf{Position} & \textbf{YoE} & \textbf{Research Area} & \textbf{LLM-General} & \textbf{LLM-Research} \\
\hline
P1  & 25-34  & Man   & PhD student & 3-5  & GenAI safety and ethics & Extensively & Occasionally \\
P2  & 25-34  & Woman & Incoming PhD student & 3-5  & HCI, AI, Accessible Computing & Frequently  & Frequently \\
P3  & 18-24  & Man   & Undergraduate student & 0-2 & NLP, Reasoning & Extensively  & Extensively \\
P4  & 25-34  & Woman & PhD student & 3-5 & HCI, AI ethics & Occasionally & Never \\
P5  & 25-34  & Woman & Master's student & 0-2 & Data Visualization & Frequently & Frequently \\
P6  & 35-44  & Man   & Postdoctoral Researcher & 3-5 & Cybersecurity & Extensively & Frequently \\
P7  & 25-34  & Man   & PhD student & 3-5 & Security and Privacy & Extensively & Frequently \\
P8  & 18-24  & Man   & Undergraduate student & 0-2  & HCI, CSCW, AR/VR, AIMC & Extensively  & Occasionally \\
P9  & 25-34  & Man   & PhD student & 6-10 & HCI, Human-AI & Extensively & Extensively \\
P10 & 25-34  & Woman & PhD student & 3-5 & HCI, Health Tracking & Frequently & Occasionally \\
P11 & 18-24  & Woman & PhD student & 0-2 & HCI, AR/VR applications & Extensively & Frequently \\
P12 & 25-34  & Woman & Master's student & 0-2 & HCI, Human-AI & Extensively  & Occasionally \\
P13 & 25-34  & Woman & PhD student & 3-5 & HCI & Rarely & Rarely \\
P14 & 25-34  & Man   & PhD student & 10+ & HCI, Metascience & Frequently & Frequently \\
P15 & 25-34  & Woman & PhD student & 6-10 & HCI, LLM/Health & Frequently & Frequently \\
P16 & 25-34  & Man   & PhD student & 3-5 & HCI, AI & Occasionally & Occasionally \\
P17 & 18-24  & Woman & Undergraduate student & 0-2 & HCI, ML & Extensively & Occasionally \\
P18 & 25-34  & Man   & PhD student & 3-5 & HCI, Social Computing & Extensively & Extensively \\
P19 & 25-34  & Woman & PhD student & 3-5 & HCI, Usable Security & Occasionally & Rarely \\
P20 & 18-24  & Woman & Incoming PhD student & 3-5 & HCI, AI, Accessibility & Frequently & Extensively \\
P21 & 25-34  & Woman & PhD student & 3-5 & HCI & Extensively & Frequently \\
P22 & 25-34  & Man   & PhD student & 3-5 & HCI, ML, UbiComp & Extensively & Extensively \\
P23 & 25-34  & Woman & PhD student & 3-5 & HCI, Emotion \& Wellbeing & Frequently  & Frequently \\
\hline
\end{tabular}
\caption{User study participants. \textbf{YoE} refers to years of experience conducting scholarly research. \textbf{LLM-General} refers to how frequently LLM applications are used for everyday tasks. \textbf{LLM-Research} refers to how frequently LLM applications are used for research activities.}
\label{tab:participantDemographics}
\end{table*}

\subsection{Additional Task Details} \label{appendix:taskDetails}
\subsubsection{Task Scenario and Instructions} The following hypothetical survey paper writing scenario was presented to participants in each task, informing their exploration over the collection of 50 papers.

\renewenvironment{quote}
  {\list{}{\leftmargin=1em \rightmargin=1em}\item\relax}
  {\endlist}
\begin{quote}
Imagine that you are part of a research team planning to write a survey paper reviewing recent research on the following topic: <topic>. You have conducted an initial search and found a set of possibly relevant papers. Today, your goal is to review this set of papers to get a broad understanding of this topic. The research team has sketched an initial outline and assigned you with two top-level sections to explore and add detail to. Using the provided papers, you could:
\begin{itemize}
\item Create one or more subsections that would be relevant to include in each section.
\item Note down specific details, e.g., terms, concepts, or perspectives, relevant to each section.
\item Cite specific papers relevant to each section, e.g., with title or author+year.
\end{itemize}
You will have up to 30 minutes to work on the outline. You are not expected to fully complete the outline or consider every paper---try to do as much as you can in the given time.
\end{quote}

\subsubsection{Creating the Sampled Paper Collections}
For each survey paper, we retrieved the full list of references using the Semantic Scholar API. Papers without open-access PDFs were excluded. To standardize the collection size, each set was limited to the 50 most recent papers---a size for which cognitive challenges of large-scale review could be observed, while still being relatively manageable with current LLM tools for creating an initial review outline.

\subsection{Post-Task Survey Questions}
Table~\ref{tab:post-task-survey} lists the post-task survey questions (7-point Likert scale) participants completed after system condition in our user evaluation.
\label{appendix:postTaskSurveyQuestions}
\begin{table*}[ht]
\centering
\small
\begin{tabular}{ll}
\toprule
\textbf{Code} & \textbf{Statement} \\
\midrule
\textbf{Reduce effort to extract information} & The system helped reduce the mental effort required to extract and organize information. \\
\textbf{Categorize papers effectively} & I was able to identify and categorize relevant papers in a meaningful way. \\
\textbf{Discover meaningful connections} & The system helped me discover connections between papers that I might have missed otherwise. \\
\textbf{Maintain control} & I maintained appropriate control over the literature review process when using this system. \\
\textbf{Verify information} & I believe I could easily verify information provided by the system. \\
\textbf{Confidence in review quality} & I feel confident in the quality and comprehensiveness of the outline I created using this system. \\
\bottomrule
\end{tabular}
\caption{Post-task survey questions and their corresponding labels (rated on 7-point Likert scale).}
\label{tab:post-task-survey}
\end{table*}


\section{Creating Paper Collections}
\begin{figure*}[ht]
    \centering
    \includegraphics[width=0.9\textwidth]{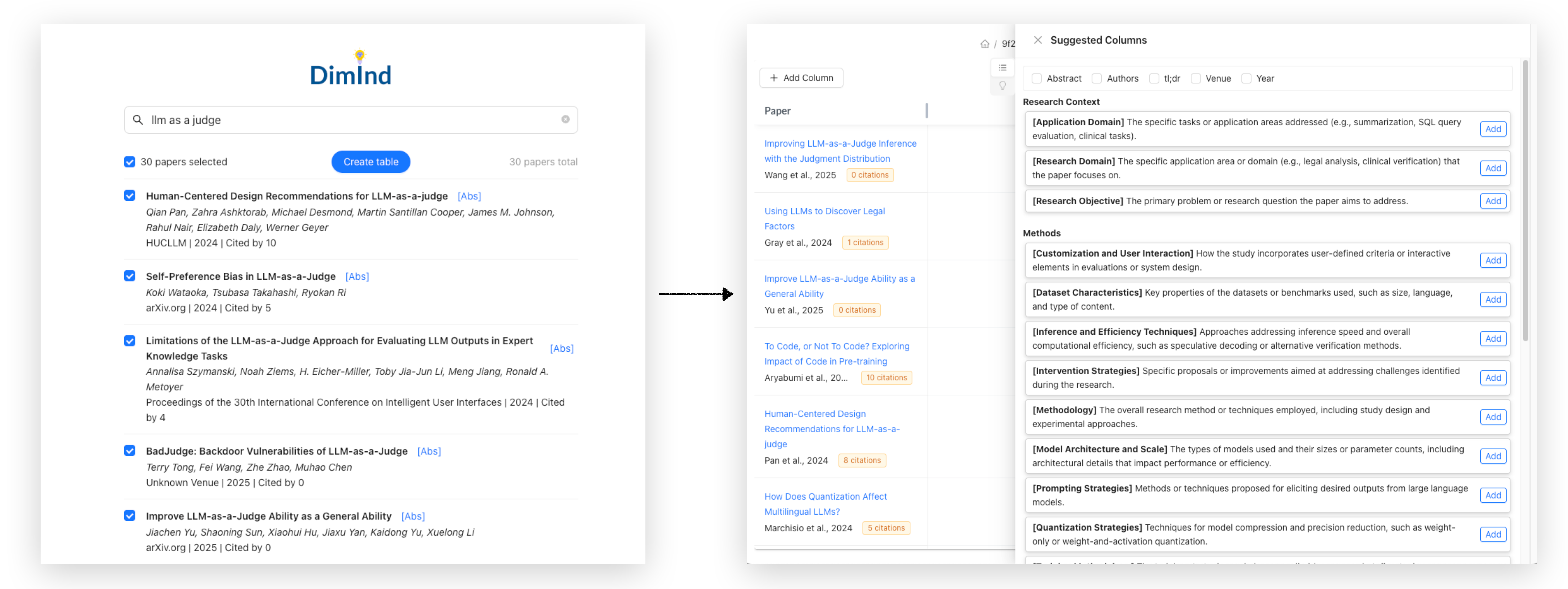}
    \caption{Paper collections in \sys{} can be created interactively from a research question or search query.}
    \Description{}
    \label{fig:table-creation-from-query}
\end{figure*}

Paper collections in \sys{} can be interactively initialized, e.g., from a research question or search query (Figure~\ref{fig:table-creation-from-query}). Finding relevant papers is a critical yet challenging part of the literature review workflow, and one that we considered out of scope for the particular exploration of \sys{} in this paper.

\clearpage

\section{\sys{} LLM Prompts} \label{appendix:prompts}

\subsection{Inducing Collection-Aware Facets}
\subsubsection{Facet Induction}
\leavevmode
\begin{framed}
\tiny
\ttfamily
\setlength{\parindent}{0pt}
A user wants to write a literature review for a set of related research papers. The following is a list of contexts from the papers. Your task is to identify facets whose values would likely allow a user to meaningfully compare and contrast across the papers.\\

Context: \{\textbf{context}\}\\

Generate facets that can be used to compare and contrast different aspects of information across the set of papers. Each facet should be a short phrase that can be used to compare information across the papers. For each facet, generate a description in the form of a short question that the facet would help answer. Generate at most \{max\_facets\} facets. The challenge is to find specific facets that are relevant to this set of papers, in addition to generally useful facets for comparing research papers, such as 'Study design' or 'Research questions'.\\

Keep each facet focused on a SINGLE concept. Do not combine multiple concepts into one facet. For example, instead of "Evaluation Metrics and Results", create separate facets for "Evaluation Metrics" and "Study Results". Instead of "Implications and Future Work", create separate facets for "Research Implications" and "Future Work Directions".\\

For example, specific facets such as 'User study methodology', 'Number of participants', or 'System design goals' could be relevant for helping a user explore a set of HCI papers. Or with a set of papers on machine learning for information retrieval, potentially informative facets may be 'Number of parameters' or 'Retrieval methods'.\\

Here are some other examples of facets: Intervention effects, Study design, Study objectives, Theoretical framework, Research questions, Dataset characteristics, Study count, Study duration, Statistical techniques, Algorithm type, Software tools, Participant demographics, Policy recommendations, Design goals, Research limitations, Ethical considerations, etc.\\

Return a single valid JSON list (without code block) containing objects with the name and description of each facet. Do not return any other text.\\

Example output:
[
    \{\{ "name": "<facet 1>", "description": "<description of facet 1>" \}\},
    \{\{ "name": "<facet 2>", "description": "<description of facet 2>" \}\},
    \{\{ "name": "<facet 3>", "description": "<description of facet 3>" \}\},
    ...
]
\end{framed}

\subsubsection{Facet Merge}
\leavevmode
\begin{framed}
\tiny
\ttfamily
\setlength{\parindent}{0pt}
The following list contains facets used to analyze research papers. Your task is to: \\
* Identify and consolidate truly duplicate facets (exact matches or synonyms) \\
* Keep distinct concepts separate, even if they appear related \\
* Select up to \{\textbf{max\_facets}\} most important facets for understanding research papers \\

Guidelines:\\
* Do not combine distinct concepts with "and" (e.g., keep "Data Collection" and "Data Analysis" as separate facets)\\
* Only consolidate facets that mean exactly the same thing (e.g., "User Study" and "User Evaluation" could become just "User Study")\\
* Each facet should represent a single clear concept\\
* Preserve specificity where it adds value\\

The following is the list of facets to process:\\
\{\textbf{facets}\}\\

Return your output as a JSON array (without code block). Each facet should have:\\
* 'name': A concise label for the facet\\
* 'type': The data type (text, number, or boolean)\\
* 'description': A clear, focused description of what the facet captures\\

Example output:
[
\{\{ "name": "Research Objective", "type": "text", "description": "The primary goal or aim of the research" \}\},
\{\{ "name": "Methodology", "type": "text", "description": "The research method used to conduct the study" \}\}
]
\end{framed}

\subsection{Extracting Faceted Information}
\subsubsection{Value Extraction}
\leavevmode
\begin{framed}
\tiny
\ttfamily
\setlength{\parindent}{0pt}
Use the provided paper context to retrieve information relevant to the specified list of facets. Each facet has a number id, a "Name", and a "Description" (optional). When provided, the "Description" key provides context (e.g., additional instructions or example output formats) for the information expected to be extracted for that facet.\\

* For each facet, generate a paragraph of detailed, accurate, and relevant information (typically between 3 to 5 sentences) by synthesizing relevant information from the provided paper context.\\
* Use a passive or third-person voice when summarizing information for an facet. For instance, avoid using phrases such as "we", "our approach", etc.\\
* If there is no relevant information for an facet in the provided context, return null for that facet value. Do not return any other text and do not make up an answer unsupported by the paper context.\\

Your output should contain a list of objects. Each object should have:\\
* 'facet\_id': The id of the facet (as provided in the input).\\
* 'value': The information extracted for that facet, or null if no relevant information is found.\\

The output should contain exactly as many objects as the facets provided in the input.\\

Paper Context:\\
\{\textbf{context}\}\\

Facets:\\
\{\textbf{facets}\}\\

Output format:\\
Do not include any explanations, only provide a valid JSON response (without code block). For example, if you are provided with 3 facets, the output should be in the following format:
[
    \{\{ "facet\_id": 1, "value": "<information for facet 1>" \}\},
    \{\{ "facet\_id": 2, "value": "<information for facet 2>" \}\},
    \{\{ "facet\_id": 3, "value": null \}\}
]
\end{framed}

\subsubsection{Value Distillation}
\leavevmode
\begin{framed}
\tiny
\ttfamily
\setlength{\parindent}{0pt}
Summarize information related to the following facet of research papers: \{\textbf{facet}\} (description (optional): \{\textbf{facet\_description}\}) from research paper excerpts. For each excerpt, return exactly one sentence of clear and concise information about this specific facet. The goal is for your summary to allow a user to more quickly understand \{\textbf{facet}\}, e.g., during an initial exploratory phase of literature review, and return to the original, longer excerpt later if they desire additional detail.\\

Guidelines for your summaries:\\
* Focus only on information directly related to \{\textbf{facet}\}\\
* Keep each summary to a maximum of 20 words\\
* Use present tense and consistent formatting across all summaries\\
* If an excerpt contains no relevant information about \{\textbf{facet}\}, return an empty string\\
* Include key statistics and metrics when present\\
* Avoid subjective interpretations or evaluations\\

Your output must be a valid JSON object (without code block) where:\\
* Keys are the original paperIds\\
* Values are either null or a string containing a single sentence summary\\
* Maintain the exact paperId format and order as provided in the input, making sure each paper in the input has a corresponding output value (null or string)\\

Example input for the facet "participant demographics":\\
\{\{
  "CorpusId:123": "The user study demographics consist of 32 trained undergraduates who have completed at least one course in computer science or statistics. A total of over 2100 responses were collected from these participants, ensuring a diverse pool of users with relevant academic backgrounds to assess the effectiveness of explanation methods.",
  "CorpusId:456": "The user study demographics include 72 participants, with a gender distribution of 19 women and 2 who declined to state their gender. Participants' ages range from 18 to over 50, with a majority falling between 30 and 39 years old. The study does not involve expert participants, but efforts were made to enhance their domain knowledge through training tasks.",
  "CorpusId:300": null
\}\}\\

Example output:
\{\{
  "CorpusId:123": "32 trained undergraduates with computer science or statistics background provided over 2100 responses.",
  "CorpusId:456": "72 participants including 19 women, aged mostly 30-39, with no expertise but received training tasks.",
  "CorpusId:300": null
\}\}\\

Now, summarize the following excerpts, returning your response as a valid JSON object (without code block):\\
\{\textbf{excerpts}\}
\end{framed}

\subsection{Synthesizing}
\subsubsection{Taxonomy Generation}
\leavevmode
\begin{framed}
\tiny
\ttfamily
\setlength{\parindent}{0pt}
A researcher is analyzing a collection of papers to write a literature review. They have extracted snippets of relevant information related to the following facet:\\
\{\textbf{facet\_name\_and\_description}\}\\

Your task is to organize these snippets into a hierarchical structure that provides clear, specific, and informative categorization. The goal is to create a nested organization that logically groups similar evidence across papers together.\\

Hierarchy Requirements:\\
* Construct a hierarchy with a maximum depth of 5 levels.\\
* Let the natural structure of the snippets determine how many levels are appropriate.\\
* The final level must contain only arrays of snippet indices (0-based indexing).\\
* The top level must have no more than \{\textbf{max\_n\_categories}\} categories.\\
* Aim to have categories that are not too small (e.g., 1 or 2 snippets) or too large (e.g., 20+ snippets). The goal is to have snippets within each category be meaningful when examined together.\\
* Use subcategories when snippets cover diverse aspects that deserve separation.\\
* Avoid vague or generic category names.\\
* Ensure that ALL snippets are included somewhere in your hierarchy.\\
* A snippet can belong to one or more categories (i.e., it may make sense for a snippet to be included in multiple relevant categories if it covers multiple topics).\\

Use "Miscellaneous" Categories Sparingly:
* Do not create a "Miscellaneous" category unless absolutely necessary.\\
Check for patterns before grouping snippets under a catch-all category.\\
Split "Miscellaneous" categories into subcategories when possible.\\
Keep them small—only use them for genuinely diverse snippets.\\

Input paper snippets:\\
\{\textbf{snippets}\}\\

Output format:\\
The output must be a valid JSON object (wihtout code block), where:
* Keys represent categories and subcategories.\\
* Values are either subcategories or arrays of snippet indices.\\
* Do not include additional keys (e.g., "indices", "description", "items").\\
* Do not include any explanation, preamble, or additional formatting.\\

Example:\\
If the input snippets were about ``performance metrics'', a good hierarchy could be:\\
\{\{
  "Model Performance": \{\{
    "Text Processing": \{\{
      "Classification": \{\{
        "Accuracy Metrics": [0, 3]
      \}\},
      "Named Entity Recognition": \{\{
        "Precision Metrics": [1]
      \}\},
      "Translation": \{\{
        "Error Analysis": [2]
      \}\}
    \}\}
  \}\},
  "System Efficiency": \{\{
    "Speed": \{\{
      "Response Time": [4, 6],
      "Processing Throughput": [7]
    \}\},
    "Resource Utilization": \{\{
      "Computational Resources": [5],
      "Memory Usage": [8],
      "GPU Performance": [9]
    \}\}
  \}\},
  "User Experience": \{\{
    "Satisfaction": [10, 12],
    "Reliability": [11, 13]
  \}\}
\}\}\\
\end{framed}

\subsubsection{Summarization}
\leavevmode
\begin{framed}
\tiny
\ttfamily
\setlength{\parindent}{0pt}
Transform the following taxonomy of organized information snippets from different research papers into a clear and concise summary that captures the key points related to the given facet. Your synthesis should be formatted as a valid JSON object with a single key "summary\_blocks" containing an array of objects, each with "header" and "content" keys.\\

Facet:\\
\{\textbf{facet\_name\_and\_description}\}\\

Taxonomy Structure and Paper Excerpts:\\
Excerpts are extracted from various research papers and provides information relevant to a more detailed aspect of the facet above:\\
\{\textbf{excerpts}\}\\

Papers to Highlight:\\
When provided, the following papers should be highlighted in your summary. Incorporate them smoothly:\\
\{\textbf{starred\_papers}\}\\

Additional Instructions:\\
* Structure your response as a valid JSON dictionary with exactly one key "summary\_blocks" containing an array of objects.\\
* Each object in the array should have two keys: "header" (corresponding to a top-level category from the taxonomy structure) and "content" (containing your synthesized text for that category).\\
* Headers should match the top-level structure provided in the taxonomy.\\
* Ensure all papers from the taxonomy are included in your summary across the different blocks.\\
* Your content should be primarily descriptive. Do not include introductory or concluding sentences. Do not generate statements that connect snippets in meaningless ways.\\
* Synthesize the key details into cohesive and insightful content blocks, covering as much of the provided information as possible.\\
* Use the provided hierarchical structure as a guide for organizing your summary blocks.\\
* Break your content into paragraphs when appropriate to improve readability.\\
* Prioritize the papers listed in the "Papers to highlight" section if any are listed, by ensuring they are incorporated into the summary. Make sure to incorporate them smoothly (e.g., by including more detail about those papers, but don't explicitly say phrases like "the highlighted papers examine..." or "the featured works on ...")\\
* Your entire summary across all blocks should be \{\textbf{length\_constraint}\}.\\
* Remember that these snippets you are summarizing come from different papers. Avoid language like "This paper" or "This system", since the summary is synthesizing across multiple papers.\\

Citing Sources:\\
* Your summary MUST cite all unique papers in the extracted snippets at least once, to the extent possible.\\
* Your citations MUST use the exact paperId format provided in the extracted snippets. The paper id could be any string (e.g., "12345", "CorpusId:12345", "URL:12345", etc.), and you should cite using the exact and entire paper id.\\
* Place citations [[paperId]] immediately after specific words, phrases, or concepts they support - not just at the end of sentences. For example: "The study found increased levels of protein X [[paperId1]] and decreased levels of enzyme Y [[paperId2]] in the treatment group."\\
* When multiple sources support the same specific point, group those citations together like [[paperId1, paperId2, paperId3]].\\

Output format:\\
\{\{
  "summary\_blocks": [
    \{\{
      "header": "First Top-Level Category",
      "content": "Your synthesized content with precise citations after specific terms [[paperId1]] or concepts [[paperId2]]. When listing multiple findings such as A [[paperId3]], B [[paperId4]], and C [[paperId5, paperId6]], citations should follow each item they support rather than appearing only at the end of sentences."
    \}\},
    \{\{
      "header": "Second Top-Level Category",
      "content": "More synthesized content with appropriate citations [[paperId7, paperId8]]."
    \}\}
  ]
\}\}
\end{framed}

\end{document}